\documentclass[iop]{emulateapj}
\usepackage{amsmath,amsopn,amsxtra,txfonts}
\usepackage{comment} 
\usepackage{natbib} 
\newcommand{\ha}{H$\alpha$}

\newcommand{\kms}{ \ifmmode{\rm km\thinspace s^{-1}}\else km\thinspace s$^{-1}$\fi}

\newcommand{\pc}{\ensuremath{ \, \mathrm{pc}}}
\newcommand{\kpc}{\ensuremath{\, \mathrm{kpc}}}
\newcommand{\cm}{\ensuremath{ \, \mathrm{cm}}}

\newcommand{\yr}{\ensuremath{ \, \mathrm{yr}}}

\newcommand{\dg}{\ifmmode{^{\circ}}\else $^{\circ}$\fi}

\newcommand{\hi}{\ion{H}{1}}

\newcommand{\lb}{\ifmmode{(\ell,b)} \else $(\ell,b)$\fi}

\newcommand{\vlsr}{\ifmmode{v_{\rm{LSR}}}\else $v_{\rm{LSR}}$\fi}
\newcommand{\av}{\ifmmode{A(V)}\else $A(V)$\fi}
\newcommand{\ebv}{\ifmmode{E(B-V)}\else $E(B-V)$\fi}
\newcommand{\iha}{\ifmmode{I_{\rm{H}\alpha}} \else $I_{\rm H \alpha}$\fi}
\newcommand{\ihb}{\ifmmode{I_{\rm{H}\beta}} \else $I_{\rm H \beta}$\fi}
\newcommand{\isii}{\ifmmode{I_{\ion{\rm{S}}{2}}} \else $I_{\rm [S \textsc{ ii}]}$\fi}
\newcommand{\inii}{\ifmmode{I_{\ion{\rm{n}}{2}}} \else $I_{\rm [N \textsc{ ii}]}$\fi}
\newcommand{\ioi}{\ifmmode{I_{\ion{\rm{o}}{1}}} \else $I_{\rm [O \textsc{ i}]}$\fi}

\newcommand{\vgeo}{\ifmmode{v_{\mathrm{geo}}} \else $v_{\mathrm{geo}}$\fi}

\shorttitle{Evidence for a Galactic Wind in the LMC}
\shortauthors{Barger et al.}

\altaffiltext{1}{Department of Physics \& Astronomy, Texas Christian University, Fort Worth, TX 76129, USA}
\altaffiltext{2}{Department of Physics, University of Notre Dame, Notre Dame, IN 46556 , USA}

\begin{document}

\author{K. A. Barger\altaffilmark{1,~2}}
\author{N. Lehner\altaffilmark{2}}
\author{J. C. Howk\altaffilmark{2}}

\title{Down-the-barrel and transverse observations of the Large Magellanic Cloud: evidence for a symmetric galactic wind on the near and far sides of the galaxy\footnotemark[*]}\footnotetext[*]{Based on observations made with the NASA/ESA \textit{Hubble Space Telescope}, obtained at the Space Telescope Science Institute, which is operated by the Association of Universities for Research in Astronomy, Inc. under NASA contract No. NAS5-26555.}

\begin{abstract}

We compare the properties of gas flows on both the near and far side of the Large Magellanic Cloud (LMC) disk using \textit{Hubble Space Telescope} UV absorption-line observations toward an AGN behind (transverse) and a star within (down-the-barrel) the LMC disk at an impact parameter of $3.2~\kpc$. We find that even in this relatively quiescent region gas flows away from the disk at speeds up to $\sim$$100~\kms$ in broad and symmetrical absorption in the low and high ions.  The symmetric absorption profiles combined with previous surveys showing  little evidence that the ejected gas returns to the LMC and provide compelling evidence that the LMC drives a global, large-scale outflow  across its disk, which is the likely result of a recent burst of star formation in the LMC. We find that the outflowing gas is multiphase, ionized by both photoionization (Si\textsc{~ii} and Si\textsc{~iii}) and collisional ionization (Si\textsc{~iv} and C\textsc{~iv}). We estimate a total mass and outflow rate to be $\ga10^7~{\rm M}_\sun$ and $\ga0.4~{\rm M}_\sun~\yr^{-1}$. Since the velocity of this large-scale outflow does not reach the LMC escape velocity,  the gas removal is likely aided by either ram-pressure stripping with the Milky Way halo or tidal interactions with the surrounding galaxies, implying that the environment of LMC-like or dwarf galaxies plays an important role in their ultimate gas starvation. Finally we reassess the mass and plausible origins of the high-velocity complex toward the LMC given its newly-determined distance that  places it in the lower Milky Way halo and sky-coverage that shows it extends well beyond the LMC disk. 

\end{abstract}


\maketitle

\section{Introduction}

The Large Magellanic Cloud (LMC) is the nearest disk galaxy at $50~\kpc$ away (see \citealt{1999ASSL..237..125W} and references therein) with a stellar and total mass of ${M}_{\rm \star}\approx3\times10^{9}~{M_\sun}$ and ${M}_{\rm total}\approx10^{10}~{M_\sun}$, respectively \citep{2009IAUS..256...81V} and a luminosity of $L=0.2~{L^*}$ \citep{2002AJ....124.2639V}. The LMC provides a closeup and relatively unobstructed view of another galaxy's interstellar medium (ISM) and of the processes influencing it. Galaxy interactions with the Small Magellanic Cloud (SMC; $M_{\rm SMC}\approx1/10~{M_{\rm LMC}}$; \citealt{2006ApJ...652.1213K} and \citealt{2010ApJ...721L..97B})---and possibly the Milky Way (MW; $M_{\rm MW}\approx100~{M_{\rm LMC}}$; e.g., \citealt{1996MNRAS.278..191G}, \citealt{2013ApJ...768..140B}, and \citealt{2014ApJ...794...59K})---have stirred the gas and stars in these galaxies, stripped material from them (e.g., \citealt{1998Natur.394..752P}), and triggered intense star formation in them (e.g., \citealt{2009AJ....138.1243H}). The stellar activity in these star-forming regions is thought to drive energetic winds and turbulance that distort the ISM (e.g., \citealt{2004ApJ...604..176S}, \citealt{2003ApJS..148..473K}, \citealt{2003MNRAS.339...87S}, \citealt{2006ApJ...646..205D}, and \citealt{2015ApJ...802...99K}). 

\begin{figure*}
\begin{center}
\includegraphics[trim=50 70 150 0,clip,scale=0.5,angle=0]{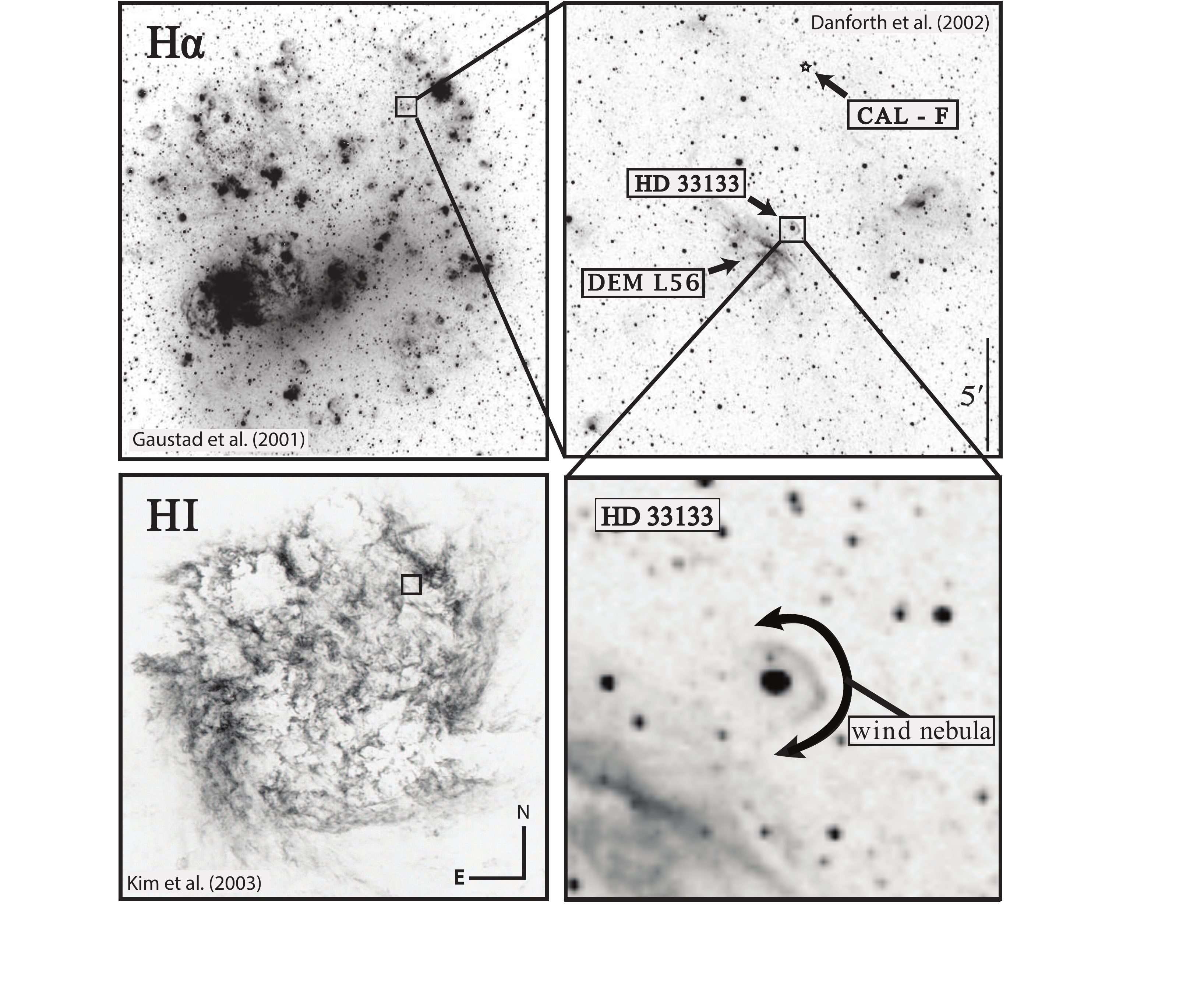}
\end{center}
\figcaption{The global \ha\ and H\textsc{~i} gas distribution of the LMC (left panels) and magnified \ha\ emission of the region surrounding the HD~33133 and CAL~F sight lines at an impact parameter of $\sim3.2~\kpc$ from the optical center of the LMC (right panels). Surrounding HD~33133 is a wind nebula produced by the strong stellar winds emanating from this WR star. To the south east of this star lies the faint H\textsc{ii} region DEM~L56. The global H\textsc{~i} and \ha\ maps are adapted from Figure~3 of \citet{2003ApJS..148..473K} using data from the Australia Telescope Compact Array (ATCA) and Parkes telescopes LMC survey and the South \ha\ Sky Survey Atlas (SHASSA: \citealt{2001PASP..113.1326G}). The right panels are adapted from images taken by \citet{2002ApJS..139...81D}.
\label{figure:ha}}
\end{figure*}

\begin{figure}
\begin{center}
\includegraphics[scale=.65,angle=0]{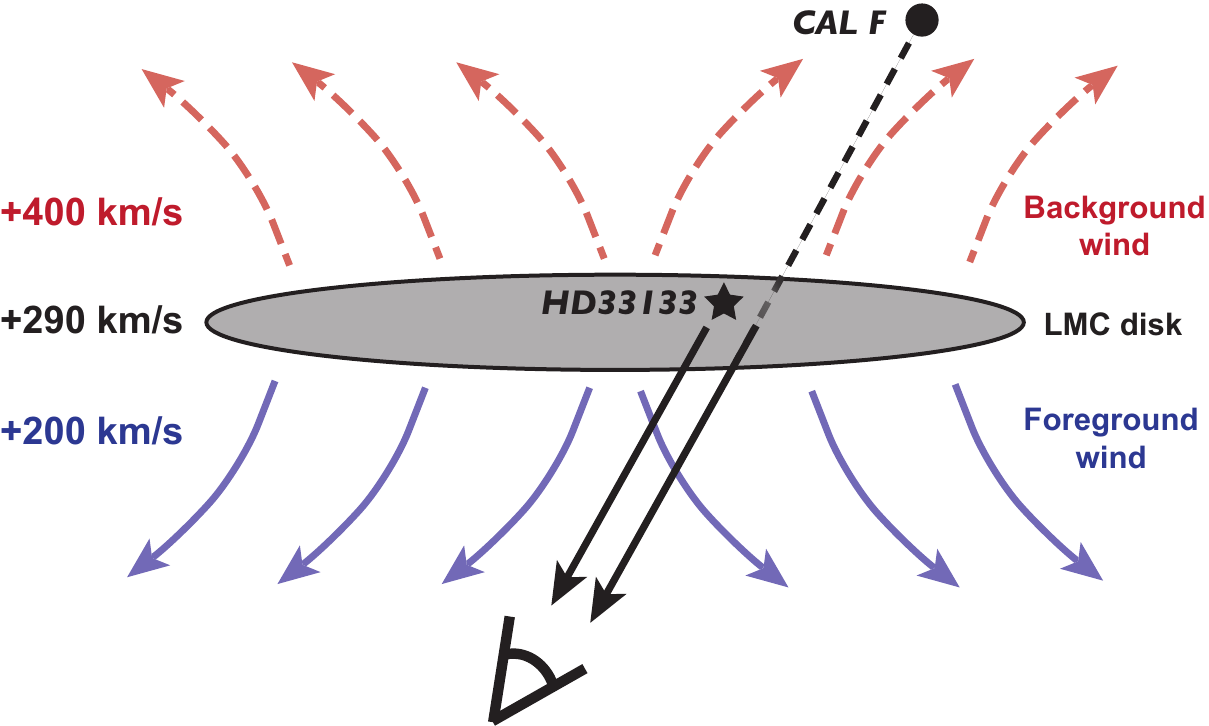}
\end{center}
\figcaption{Schematic of the line-of-sight path to the star HD~33133 positioned within the LMC disk and the background AGN CAL~F. Solid lines are used for the regions in front of the galaxy and dashed lines for the regions behind. Toward both sight lines, we detect neutral and ionized gas that extends much further than the H\textsc{~i} velocity extent of the LMC disk. The kinematics and properties of the absorption along these sight lines are consistent with material flowing away from both sides of the galaxy's disk (see Section~\ref{section:properties} below). The colored arrows illustrate the blue- and red-shifted gas distribution about the LMC in a large-scale galactic wind scenario. 
\label{figure:schematic}}
\end{figure}

Absorption line studies utilizing \textit{Hubble Space Telescope (HST)} and the \textit{Far Ultraviolet Spectroscopic Explorer (FUSE)} show gas with velocities consistent with material flowing away from this galaxy toward stars embedded within the LMC disk (\citealt{2002ApJ...569..214H} and \citealt{2007MNRAS.377..687L}, and \citealt{2011MNRAS.412.1105P}). These absorption lines probe only the material foreground to the background stars. If a large-scale outflow is being driven from the LMC, gas should flow from both the near and far sides of the galaxy. We  now have probes of gas flowing out of the backside of the LMC. To test if there is an outflow emanating from both sides of the LMC, in this paper we compare the absorbing material on the near and far side of this galaxy's disk along two neighboring sight lines as shown in Figure~\ref{figure:ha}. Located behind the LMC, the sight line to the AGN CAL~F penetrates through the entire disk, probing the regions in front, within, and behind the LMC. Embedded within the disk, the sight line to the star HD~33133 is only sensitive to the material in its foreground. Figure~\ref{figure:schematic} illustrates the line-of-sight path to these targets. The absorption along these sight lines enable us to differentiate between the gaseous flows on both sides of the disk, since they lie only $7.2\arcmin$ apart (a projected distance of $105~{\rm pc}$). These sight lines essentially probe the same environment within the disk (Figures~\ref{figure:ha}--\ref{figure:schematic}). 

Our experiment is analogous to combining the ``down-the-barrel'' galaxy absorption experiments, which probe only gas in the foregrounds of the integrated light from distant galaxies (e.g., \citealt{2009ApJ...692..187W}, \citealt{2010ApJ...719.1503R, 2014ApJ...794..156R}, \citealt{2012ApJ...759...26E}), and QSO absorption line experiments, which probe gas in both the foreground and background of the galaxy (e.g., \citealt{2011Sci...334..948T}, \citealt{2011ApJ...733..105K}, \citealt{2013ApJ...763..148S}, \citealt{2009ApJS..182..378W}, \citealt{2015ApJ...804...79L}).  In our case, our foreground probe is a single star, while the foreground and background probe is an AGN at an impact parameter of $\sim3.2~\kpc$ from the optical center of the LMC ($2.5~\kpc$ from the kinematic center; \citealt{1999AJ....118.2797K}). \citet{2014ApJ...792L..12K} have also presented a joint comparison of the foreground absorption toward a galaxy with the foreground and background absorption from the same galaxy as seen toward a QSO, albeit in this case projected $58~\kpc$ from the galaxy. In our experiment, the AGN lies behind the optical disk of the LMC (similar to the QSO experiment of \citealt{2005ApJ...635..880B}).

The HD~33133 and CAL~F sight lines investigated in this study probe a relatively quiescent region of the galaxy known as the ``Blue~Arm" that has a star-formation rate (SFR) of ${\rm SFR}_{\rm Blue~Arm}\lesssim0.03~M_{\sun}~\yr^{-1}$  compared to the LMC's total rate of ${\rm SFR}_{\rm LMC}\lesssim0.2~M_{\sun}~\yr^{-1}$ (or ${\rm sSFR}=2\times10^{-11}~\yr^{-1}$) (\citealt{2009AJ....138.1243H} and \citealt{2009IAUS..256...81V}), allowing us to determine if the outflows observed in the \citet{2002ApJ...569..214H}, \citet{2007MNRAS.377..687L}, and \citet{2011MNRAS.412.1105P} studies persists throughout the LMC. However, it is important to note that although the Blue~Arm is much less active today, it did undergo a major star formation event 100--160$~{\rm Myr}$ ago that formed the equivalent of five times the mass of the 30~Doradus' stellar population \citep{2009AJ....138.1243H}.

This paper is organized as follows. In Section \ref{section:observations}, we discuss the UV spectroscopic absorption observations and their reduction. We examine the properties of the absorbing material in Section \ref{section:properties}. We discuss the potential causes for the broad absorption that extends roughly $100~\kms$ off the \hi\ disk along the HD~33133 and CAL~F sight lines in Section~\ref{section:discussion}, which includes a rotating thick disk, tidal debris, and a large-scale, feedback-driven wind. Finally, we conclude with our major results in Section \ref{section:summary}.  

\clearpage

\begin{figure*}
\begin{center}
\includegraphics[scale=0.65,angle=0]{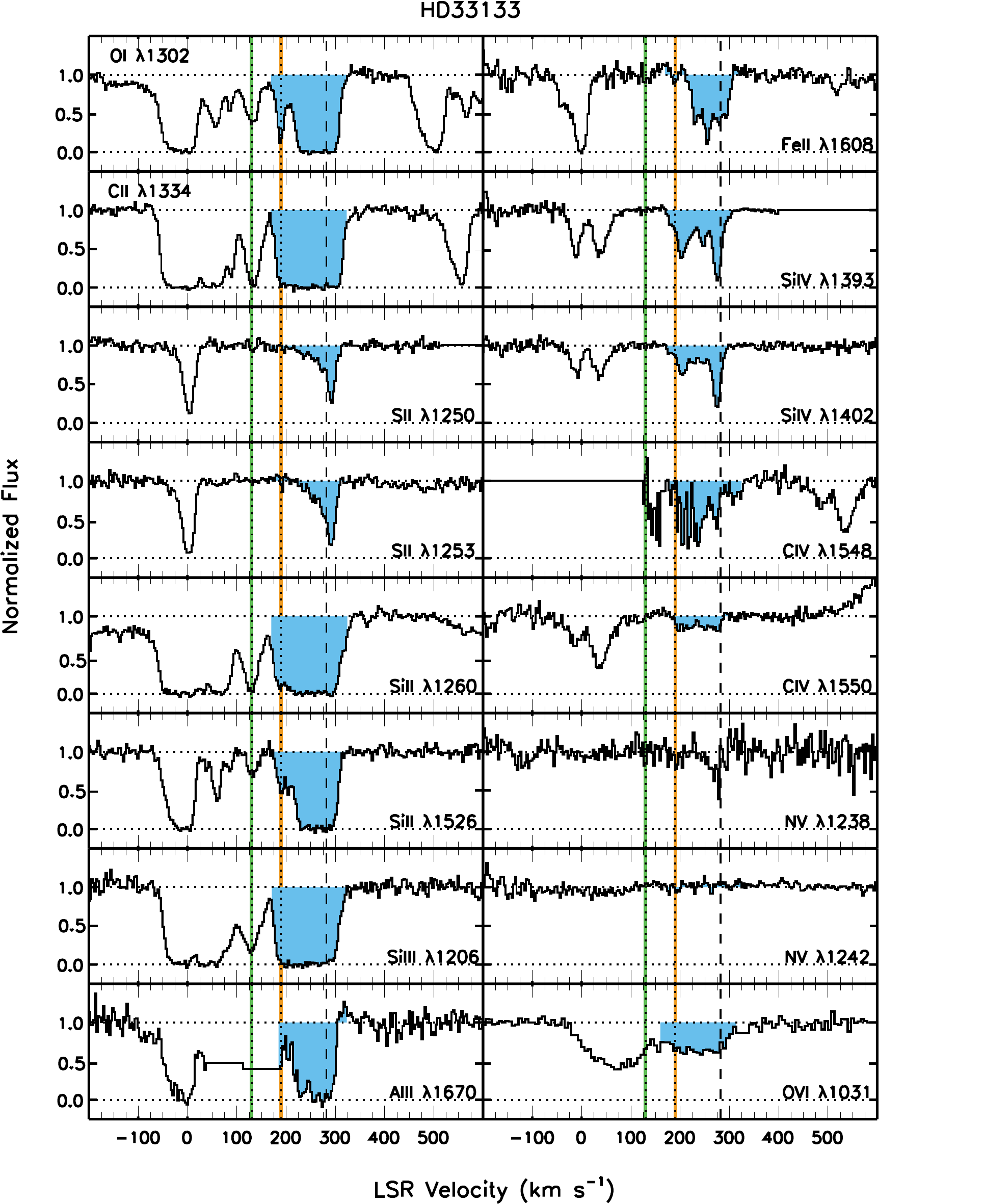}
\end{center}
\figcaption{Normalized flux of HD~33133 as a function of the LSR velocity. The vertical dashed line at $\vlsr=+282~\kms$ indicates the peak velocity of the LMC HI emission, as illustrated in Figure~\ref{figure:hi}, and the blue shaded region marks the extent of kinematically continuous gas within and around it from $+170\lesssim\vlsr\lesssim+325~\kms$. We indicate with the green vertical line an HVC at $\vlsr=+130~\kms$  that is part of the high-velocity cloud (HVC) complex toward the LMC (see Section~\ref{section:HVC} and \citealt{2009ApJ...702..940L}) and the orange vertical line at $\vlsr=+190~\kms$ marks the wind nebula surrounding this star (see Figure~\ref{figure:ha}). The absorption at $\vlsr\lesssim90~\kms$ coincides with the disk and halo of the MW. The ${\rm C}{~\textsc{ii}*}~\lambda1335$ absorption of the MW is blended with the ${\rm C}{~\textsc{ii}}~\lambda1334$ transition of the LMC. The feature seen at $\vlsr\approx+275~\kms$ in the ${\rm N}{~\textsc{v}}~\lambda1238$ panel is MW absorption from ${\rm Mg}{~\textsc{ii}}~\lambda1239$. All of these spectra were acquired with \textit{HST}/STIS except the ${\rm O}\textsc{vi}~\lambda1031$ in the bottom-right panel, which was obtained with \textit{FUSE}. 
\label{figure:HD33133_norm}}
\end{figure*}

\begin{figure*}
\begin{center}
\includegraphics[scale=0.65,angle=0]{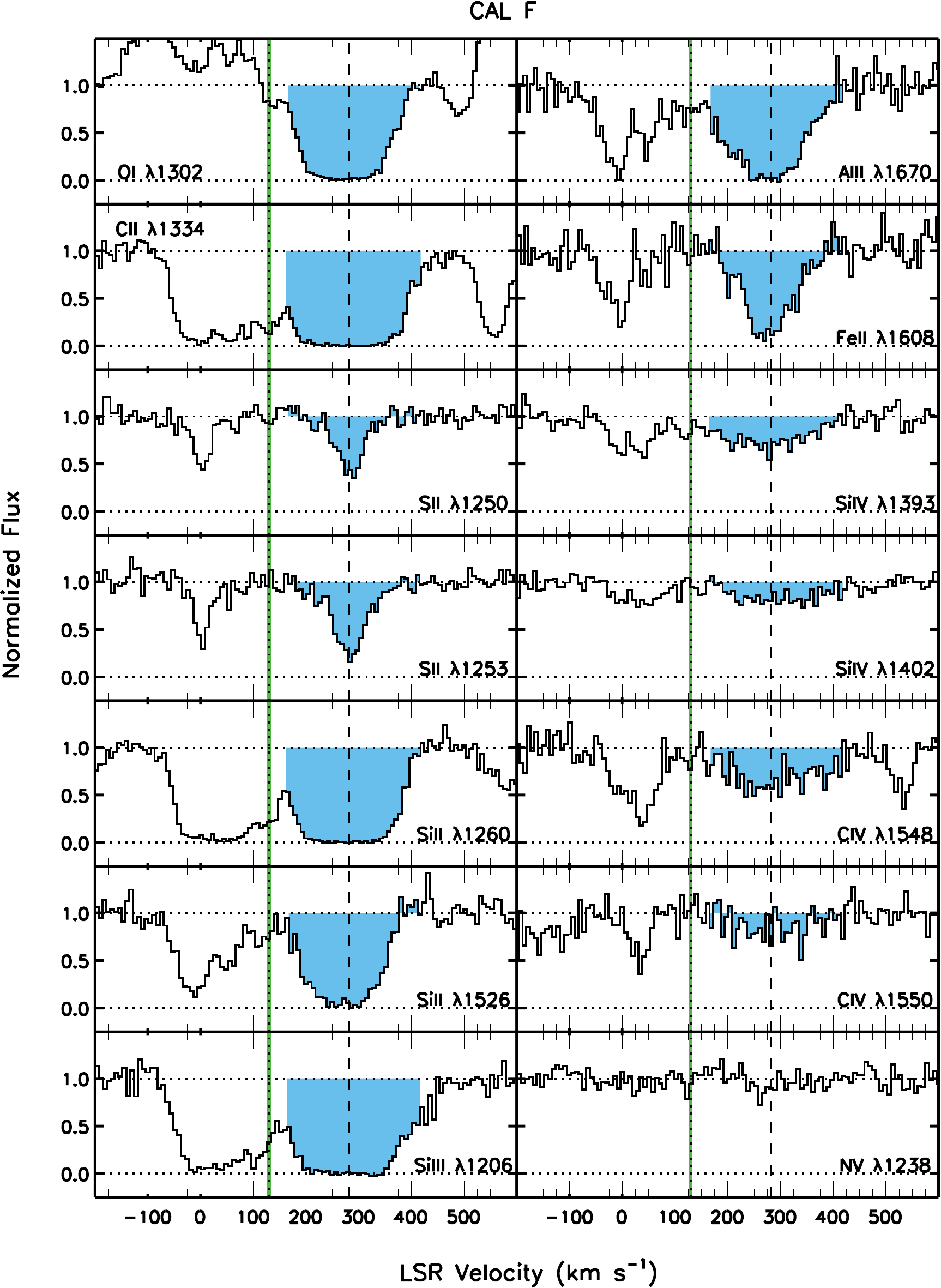}
\end{center}
\figcaption{Normalized flux of CAL~F as a function of the LSR velocity. The vertical dashed line at $\vlsr=+282~\kms$ indicates the peak velocity of the LMC HI emission, as illustrated in Figure~\ref{figure:hi}, and the blue shaded region marks the extent of kinematically continuous gas within and around it from $+165\lesssim\vlsr\lesssim+415~\kms$. The green vertical line at $\vlsr=+130~\kms$ marks an HVC complex in the direction of the LMC in the ${\rm C}\textsc{~ii}~\lambda1334$, ${\rm Si}\textsc{~ii}~\lambda1260,~1526$, and ${\rm Si}\textsc{~iii}~\lambda1206$ ions (the feature in ${\rm S}\textsc{~ii}~\lambda1250$ is unrelated as it is not observed in ${\rm S}\textsc{~ii}~\lambda1253$). The absorption at $\vlsr\lesssim90~\kms$ coincides with the disk and halo of the MW. The ${\rm C}{~\textsc{ii}*}~\lambda1335$ absorption of the MW is blended with the ${\rm C}{~\textsc{ii}}~\lambda1334$ transition of the LMC. The $\rm O\textsc{~i}~\lambda1302$ is contaminated by strong O\textsc{~i} airglow emission lines near $\vlsr=-100,~0,~{\rm and}~+550~\kms$. The feature seen at $\vlsr\approx+275~\kms$ in the ${\rm N}{~\textsc{v}}~\lambda1238$ panel is MW absorption from ${\rm Mg}{~\textsc{ii}}~\lambda1239$. The LMC absorption along this sight line extends beyond $\vlsr\approx+325~\kms$ absorption detected along the LMC HD~33133 disk star sight line (see Figure~\ref{figure:HD33133_norm}).
\label{figure:cal_f_norm}}
\end{figure*}

\section{Observations, Reduction, and Analysis} \label{section:observations}

\begin{deluxetable*}{cccccc}
\tabletypesize{\footnotesize}
\tablecolumns{4}
\tablecaption{Summary of Targets\label{table:details}}
\tablehead{
 \colhead{Target} & \colhead{Alias} & \colhead{Type} &\colhead{$({\rm R.A.,~Dec})_{\rm J2000}$}} 
 \startdata
HD~33133 &  Sk--$66\arcdeg51$, ${\rm Br}~13$ & WN8h star & $(5^{\rm h}3^{\rm m}8.84^{\rm s},-66\arcdeg40\arcmin57.30\arcsec)$ \\
CAL~F & ${\rm RX~J}0503.1{\rm-}6634$ & Seyfert~1 ($z=0.064$) & $(5^{\rm h}3^{\rm m}3.93^{\rm s},-66\arcdeg33\arcmin46.50\arcsec)$
\enddata
\end{deluxetable*}

To determine the origin of the gas flowing away from the LMC, we examine the spectroscopic differences along two neighboring sight lines: an early-type star in the disk of the LMC and a background AGN. The details of the stellar HD~33133 and AGN CAL~F targets are listed in Table~\ref{table:details} and their positions in the disk is shown in Figure~\ref{figure:ha}.

CAL\,F was observed with COS G130M and G160M and the star HD~33133 with STIS E140M (through the square aperture to maximize throughput) with a wavelength coverage of 11340-1796 \AA\ for COS and $1170-1730~\AA$ for STIS. In these wavelength ranges, some or all of the following species are covered: O\textsc{~i}~$\lambda$1302, C\textsc{~ii}~$\lambda$1334, C\textsc{~iv}~$\lambda\lambda$1548, 1550, N\textsc{~v}~$\lambda$$\lambda$1238, 1242, S\textsc{~ii}~$\lambda\lambda$1250, 1253, 1259, Si\textsc{~ii}~$\lambda$1190, 1193, 1260, 1304, 1526, Si\textsc{~iii}~$\lambda$1206, Si\textsc{~iv}~$\lambda\lambda$1393, 1402, Al\textsc{~ii}~$\lambda$1670, Fe\textsc{~ii}~$\lambda$1608.  In Table~2, we provide the observation summary for our program. We also use the {\it FUSE} observations of HD~33133 to explore the O\textsc{~vi}\ $\lambda$1031 absorption; we refer the reader to \citet{2002ApJ...569..214H} for the data reduction of the  {\it FUSE} data and the removal of the H$_2$ contamination from the O\textsc{~vi} absorption. 

Information about COS and COS data reduction can be found in \citet{2009Ap&SS.320..181F}, \citet{2012ApJ...744...60G}, and \citet{2012cosi.book.....H}. Information about STIS can be found in the STIS {\it HST}\ Instrument Handbook \citep{2011stis.book.....B}. Standard reduction and calibration procedures were used to reduce the STIS and COS data. The alignment of the individual spectra was achieved through a cross-correlation technique in wavelength space. The individual exposures (in flux units), gratings, and echelon orders were combined to produce a single spectrum. Each coadded spectrum was then shifted into the LSR velocity frame.\footnote{We use the kinematic definition of the LSR, where the solar motion moves at $20~\kms$ toward $(R.A.,~Dec)_{J2000}=(18^{\rm h}3^{\rm m}50.29^{\rm s}, +30\arcdeg00\arcmin16.8\arcsec)$ throughout this study for both the absorption- and emission-line datasets.}  As the COS absolute wavelength calibration is somewhat uncertain (up to $15-20~{\rm km}$ uncertainty), we made sure that the centroids of the Galactic \hi\ 21-cm emission in that direction from the LAB survey \citep{2005A&A...440..775K} match those seen in the Galactic absorption of neutral and singly ionized species. 

We then normalized the AGN and stellar continua with Legendre polynomials within roughly $\pm1000~\kms$ from the absorption under consideration. For the AGN, we used typically Legendre polynomials with orders of $m \le 3$. For the stellar continuum of HD~33133, we used values of $m$ ranging between 2--8 depending on the continuum complexity of the studied transitions. In Figures~3 and 4, we show the normalized profiles of several species in different ionization levels. The LMC component studied in this work is highlighted by the grey shaded area.  Note that strong O\textsc{~i} airglow emission lines contaminate part of the O\textsc{~i}~$\lambda$1302 and Si\textsc{~ii}~$\lambda$1304 absorption in the COS spectrum owing to its large aperture, but the LMC O\textsc{~i}~$\lambda$1302 component appears to be free of any contamination. 

\begin{deluxetable*}{cccccccc}
\tabletypesize{\footnotesize}
\tablecolumns{10}
\tablecaption{Summary of Observations\tablenotemark{a}\label{table:observations}}
\tablehead{
 \colhead{Target} & \colhead{Instrument} & \colhead{Aperture} & \colhead{Grating} &\colhead{$\lambda_{\rm center}$} &\colhead{$\lambda$~range} & \colhead{Resolution} & \colhead{Total~$t_{\rm exp}$} \\ 
 \multicolumn{4}{c}{} & \colhead{($\AA$)} & \colhead{($\AA$)} & \colhead{($\kms$)} & \colhead{(ks)} }  
 \startdata
HD~33133 &  STIS & $0.2\arcsec\times0.2\arcsec$ & E140M  & 1425 & 1140--1735 & $6.5$ & 2.8\\
CAL~F & COS & $2.4\arcsec$ & G130M & 1291 & 1140--1430 & $17$  & 4.7 \\
CAL~F & COS & $2.4\arcsec$ & G160M & 1600 & 1385--1750 & $17$  & 4.0
\enddata
\tablenotetext{a}{\textit{HST} program ID~11692 (PI: Howk).}
\end{deluxetable*}

We use the apparent optical depth (AOD) method---described in \citet{1991ApJ...379..245S}---to convert the observed absorption to column densities. This method relates the observed flux~($F_{obs}$), the flux of the continuum~($F_c$), and the line strength~($f\lambda_0$) of ionic transitions to determine their apparent column density without prior knowledge of their component structure:
\begin{equation}\label{eq:Na}
N_a(v)=\frac{3.768\times10^{14}}{f\lambda_0}~\ln{\left(\frac{F_c(v)}{F_{obs}(v)}\right)}~\frac{\cm^{-2}}{\kms}, 
\end{equation}
where $f$ is the oscillator strength and $\lambda_0$ is the rest wavelength of the line in units of $\AA$. We use the atomic parameters listed in \citet{2003ApJS..149..205M} for the UV transitions. 

\section{Observed Properties of the LMC Gas}\label{section:properties}

\subsection{H\textsc{~i} Emission and Systemic Velocity toward CAL~F and HD~33133}\label{section:vel_diff}

\begin{figure}
\begin{center}
\includegraphics[scale=0.35,angle=90]{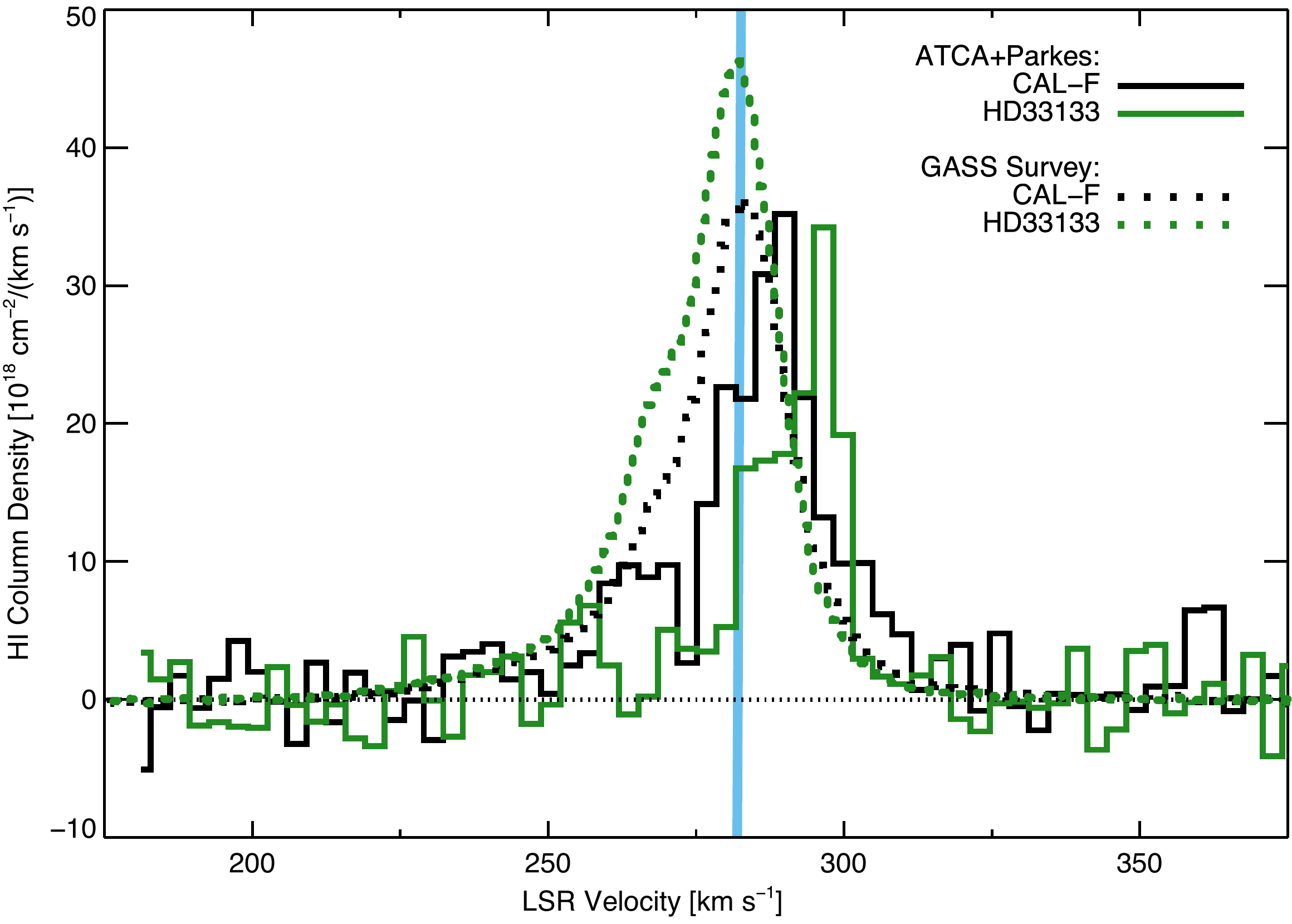}
\end{center}
\figcaption{The H\textsc{~i} gas distribution along the CAL~F (black) and HD~33133 (green) sight lines. The LMC H\textsc{~i} emission peaks at $+288~\kms$ along the CAL~F sight line and at $+296~\kms$ towards HD~33133 when resolved at $1\arcmin$ (ATCA and Parkes survey; \citealt{1998ApJ...503..674K, 2003ApJS..148..473K, 2003MNRAS.339...87S}). The H\textsc{~i} of both sight lines peaks at $+282~\kms$ (blue line) when resolved over a larger $16\arcmin$ region (Parkes Galactic All-Sky Survey;  \citealt{2009ApJS..181..398M} \& \citealt{2010AA...521A..17K}). 
\label{figure:hi}}
\end{figure}

\begin{figure*}
\begin{center}
\includegraphics[trim=20 61 0 0,clip, scale=0.515,angle=0]{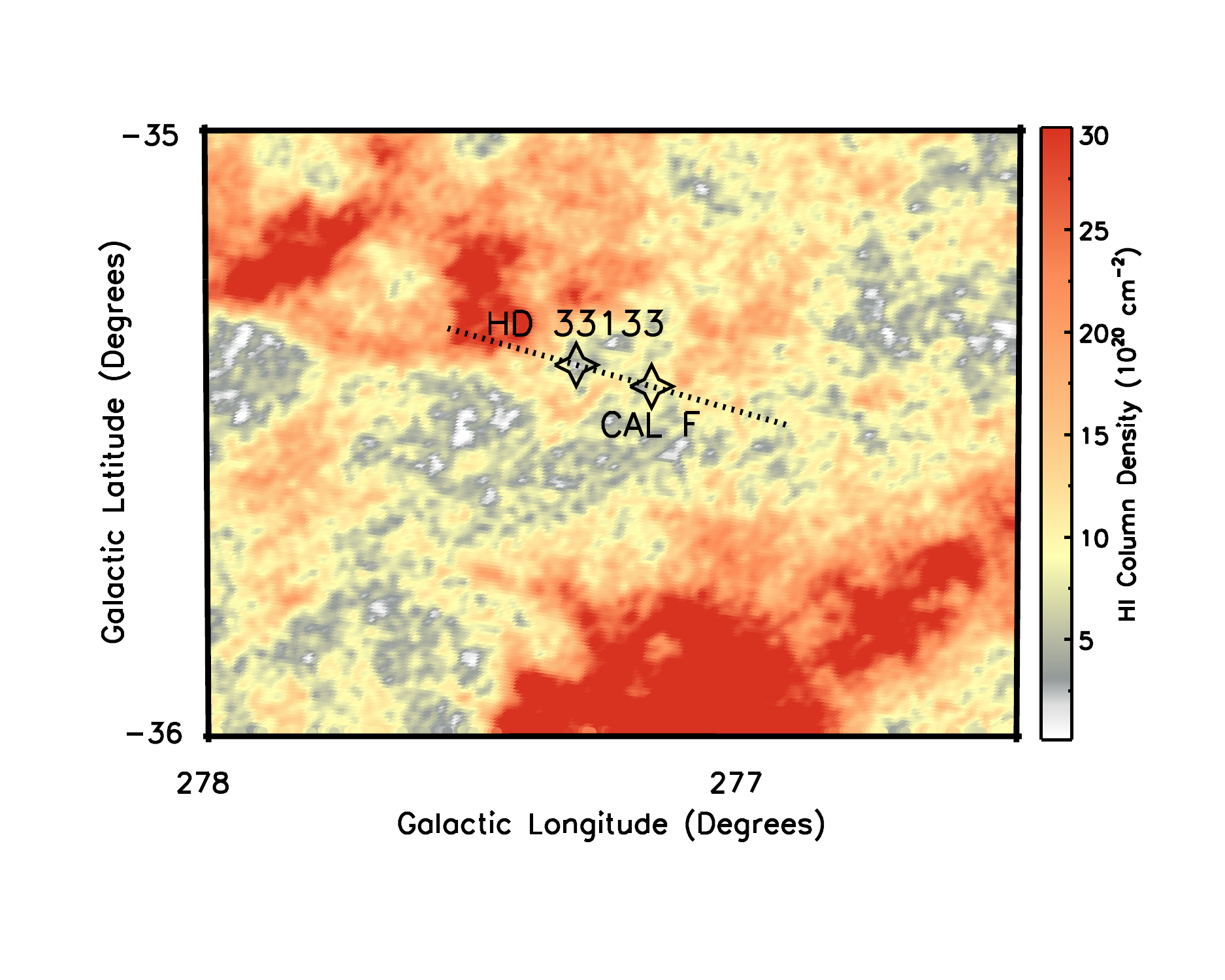}
\includegraphics[trim=0 -15 0 25,clip, scale=0.4,angle=0]{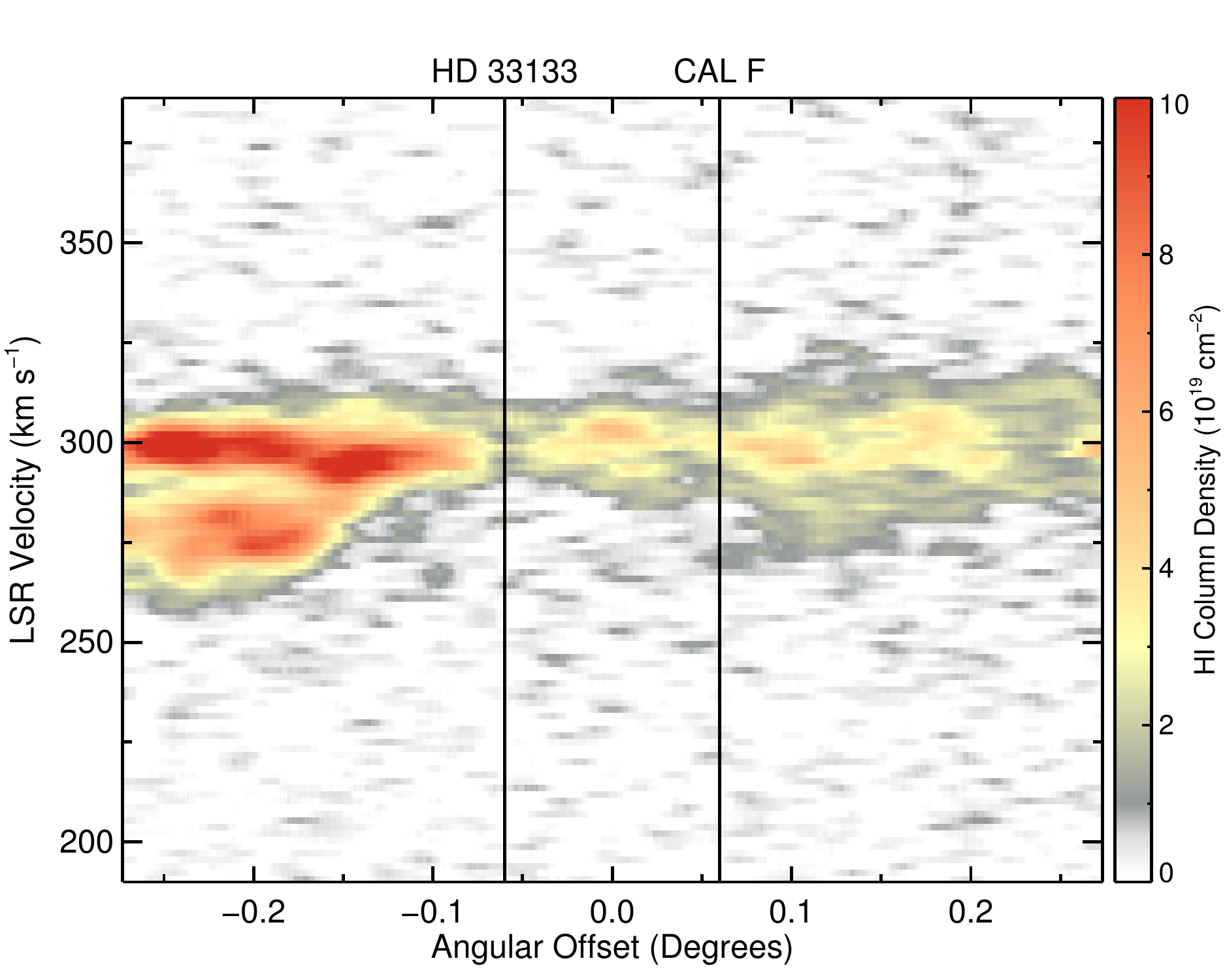}
\end{center}
\figcaption{H\textsc{~i} emission maps of the gas surrounding the HD~33133 and CAL~F sight lines. The left figure shows a map of the H\textsc{~i} emission integrated over the $+190\le\vlsr\le+385~\kms$ range---the full range of the combined ATCA H\textsc{i} and Parkes telescopes LMC survey (\citealt{1998ApJ...503..674K} and \citealt{2003ApJS..148..473K}). The right map shows a position-velocity diagram along the dotted line in the integrated H\textsc{~i} map. 
\label{figure:hi_distribution}}
\end{figure*}

The \hi\ emission in the region surrounding the stellar HD~33133 and AGN CAL~F sight lines spans $+250\lesssim\vlsr\lesssim+310~\kms$ with the emission peaking near $\vlsr\approx+282~\kms$ (see Figures~\ref{figure:hi} and~\ref{figure:hi_distribution}). We use the \hi\ emission from the LMC to define the velocity extent of the neutral gas in its disk. The distribution of \hi\ emission along these two sight lines is very similar with their peaks offset by only $\sim8~\kms$ (see Figure~\ref{figure:hi}, combined ATCA and Parkes \hi\ observations). Unlike the absorption toward LMC disk stars, this emission traces the entire extent and kinematics of the disk, though it is less sensitive to the gas at the lowest densities than absorption-line studies. We adopt $\vlsr\approx+282~\kms$ as the systemic velocity of the LMC for this region of the galaxy and the velocity spread of the \hi\ emission as a reference for the extent of the gaseous disk when identifying LMC intermediate-velocity clouds (IVCs) that span up to $100~\kms$ from the \hi\ emission of LMC disk and the HVCs that span to even greater velocities. Although the ionized  thick disk can extend over a wider range of velocities, the bulk of the neutral gas in the LMC disk resides at $+250\le\vlsr\le+310~\kms$ (Figure~\ref{figure:hi}). 

\subsection{Comparison of the Kinematic Structure between CAL~F and HD~33133}\label{section:vel_collisional}

The stellar HD~33133 and AGN CAL~F sight lines have absorption features at $-50\lesssim\vlsr\lesssim+100~\kms$ that trace the MW and absorption features at $\vlsr\approx+130~\kms$ that traces an HVC that covers the face of the LMC (\citealt{2003MNRAS.339...87S} and \citealt{2009ApJ...702..940L}) in common (see Figures~\ref{figure:HD33133_norm} and~\ref{figure:cal_f_norm}). The absorption at $\vlsr\approx+130~\kms$ is less prominent along the CAL~F sight line as the COS spectrograph has a velocity resolution that is $\sim2.6\times$ higher than the STIS spectrograph, though this absorption is still present in the ${\rm C}\textsc{~ii}~\lambda1334$, ${\rm Si}\textsc{~ii}~\lambda1260,~1526$, and ${\rm Si}\textsc{~iii}~\lambda1206$ ions. From the observed properties of this HVC in \hi\ emission \citep{2003MNRAS.339...87S} and in absorption \citep{2009ApJ...702..940L}, this complex was likely produced through LMC outflow events as discussed later in Section~\ref{section:HVC} and in those studies. 

Toward both sight lines, continuous absorption of the LMC extends roughly $-100~\kms$ further than the \hi\ disk emission to $\vlsr\approx+175~\kms$ (see Figure~\ref{figure:resolution}). Because the HD~33133 sight line is only sensitive to material in its foreground (see Figure~\ref{figure:schematic}), this absorbing material must be moving away from the near side of the LMC. Along the CAL~F sight line, we detect extra absorption that is not seen toward HD~33133, which extends $+90~\kms$ further than the H I emission to $\vlsr\approx+415~\kms$, in neutral, low- and high-ionization species (see Figures~\ref{figure:cal_f_norm} and~\ref{figure:resolution}). Since this gas is not detected along the HD~33133 sight line, it must lie behind and be moving away from the LMC. In the following sections, we will focus on the intermediate-velocity absorption that spans up to roughly $\pm100~\kms$ off of the LMC. We discuss the LMC HVC and its origin in Section~\ref{section:HVC}.

The absorption along the stellar HD~33133 sight line has an extra narrow component near $v_{\rm LMC}\approx-90~\kms$ ($\vlsr\approx+190~\kms$; Figures~\ref{figure:HD33133_norm} and~\ref{figure:resolution}), as seen in O\textsc{~i}, Si\textsc{~ii}, Si\textsc{~iv}, and other ions, that coincides with an approaching shell wall of the wind nebula that surrounds this star as seen in the bottom-right panel of Figure~\ref{figure:ha} \citep{1999AJ....117.1433C}. In contrast, the Si\textsc{~iv} and C\textsc{~iv} absorption along the AGN CAL~F sight line only consists of a broad absorption (see Figure~\ref{figure:resolution}). Although the AGN and stellar spectra have different resolutions (see Table~\ref{table:observations}), this does not drive the disparity in their velocity structure. As displayed in Figure~\ref{figure:resolution}, the STIS spectrum still shows the extra narrower component when smoothed and rebinned to the COS resolution and sampling. These narrow components indicate that the gas near the WR star HD~33133 is strongly affected by its presence. We discuss these components in more detail in the Appendix, but we emphasize their absence in the spectrum of CAL~F, demonstrating that the region near CAL~F is does not probe an H\textsc{~ii} or supershell region (see Figure~\ref{figure:resolution}). 

\begin{figure}
\begin{center}
\includegraphics[scale=0.475,angle=0]{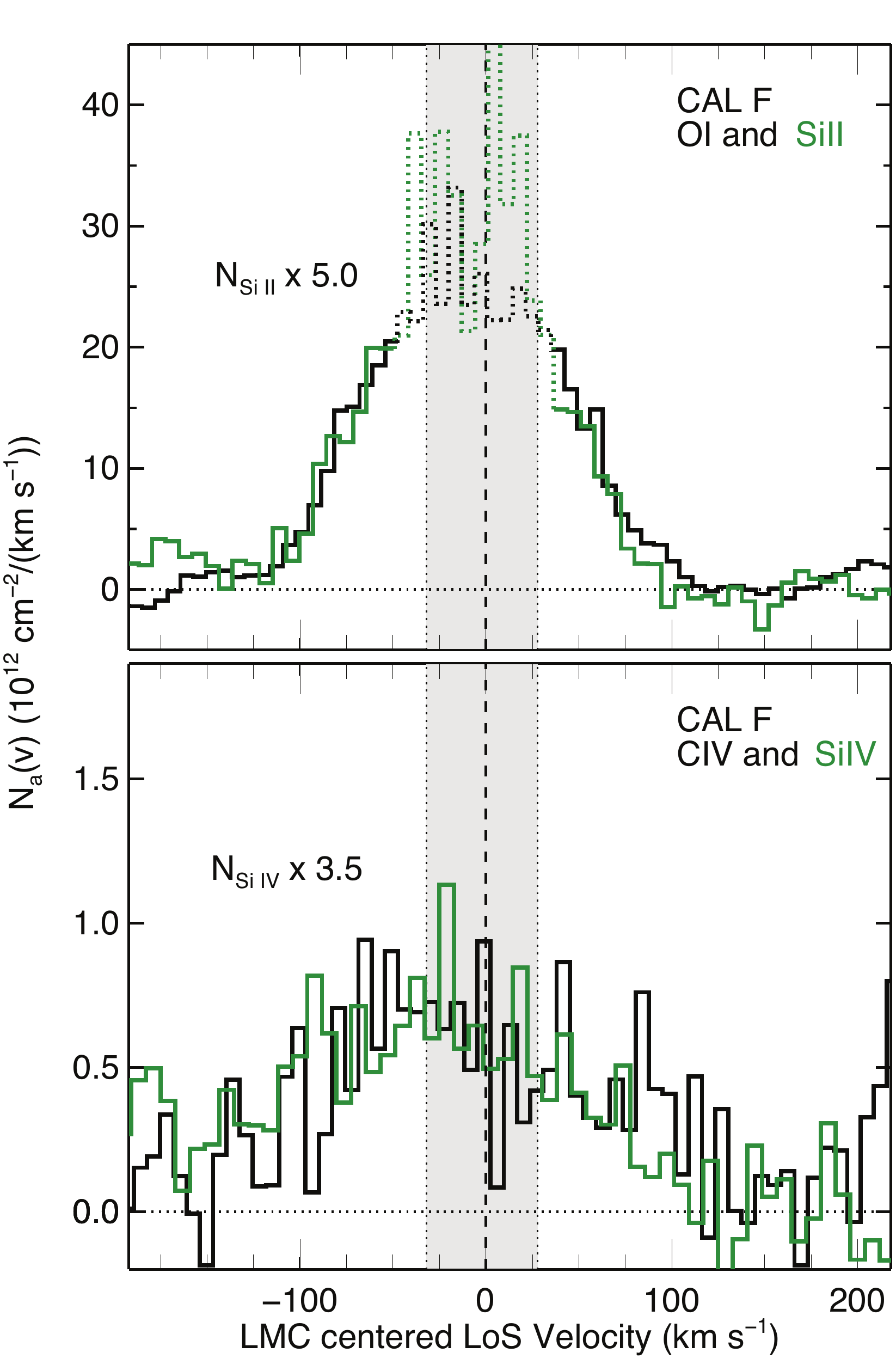}
\end{center}
\figcaption{The apparent column densities of the low (top) and high (bottom) ionization species along the AGN CAL~F sight line. These distributions are traced in green for the Si\textsc{~ii} and Si\textsc{~iv} ions and in black for the O\textsc{~i} and C\textsc{~iv} ions. The dashed portions of the spectra are regions where the absorption is saturated. The extent of H\textsc{~i} emission from the LMC disk is highlighted in light grey.   
\label{figure:OI_SiII_CIV_SiIV}}
\end{figure}

\begin{figure*}
\begin{center}
\includegraphics[scale=0.425,angle=0]{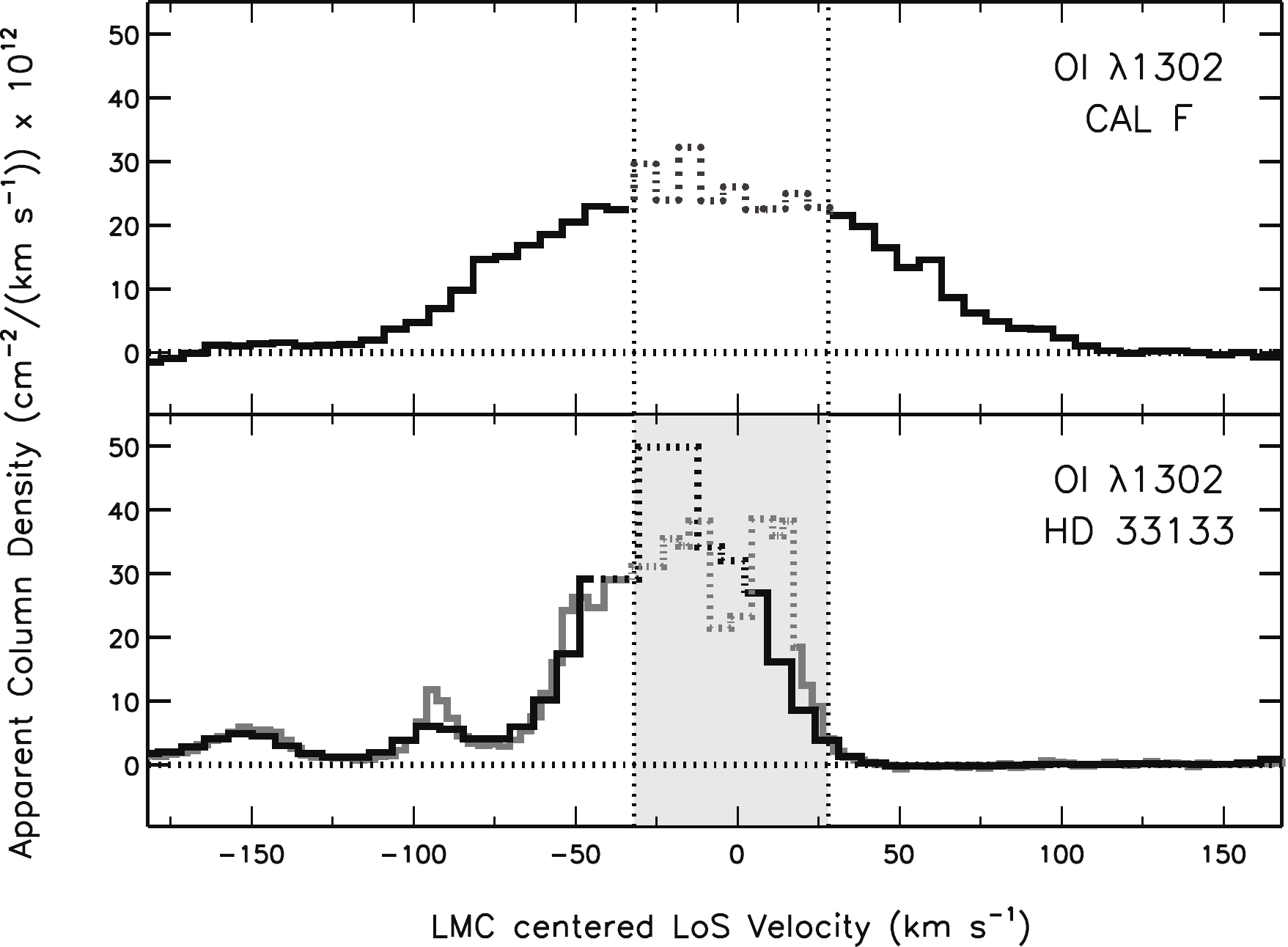}
\includegraphics[scale=0.425,angle=0]{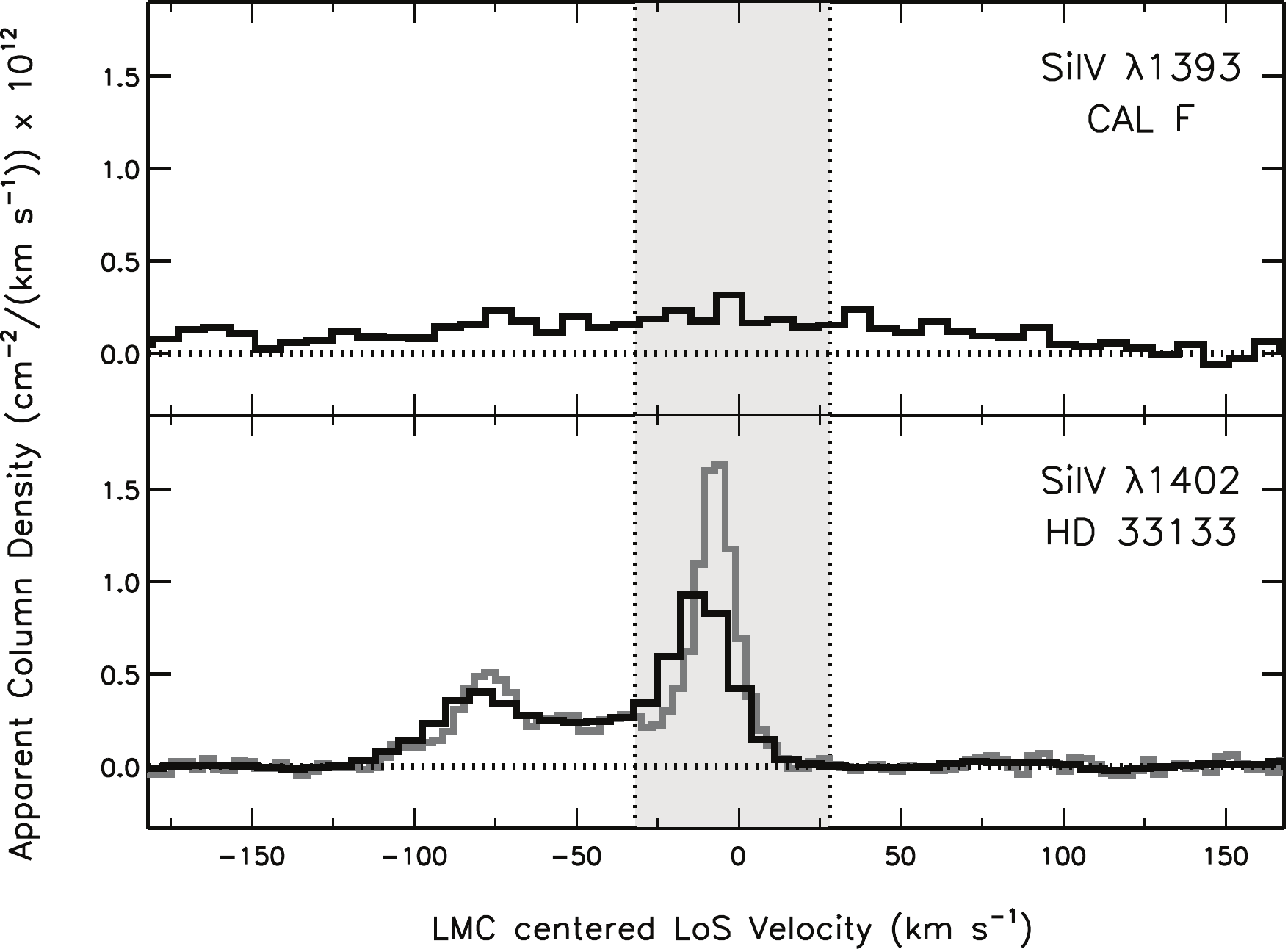} \\ 
\includegraphics[scale=0.425,angle=0]{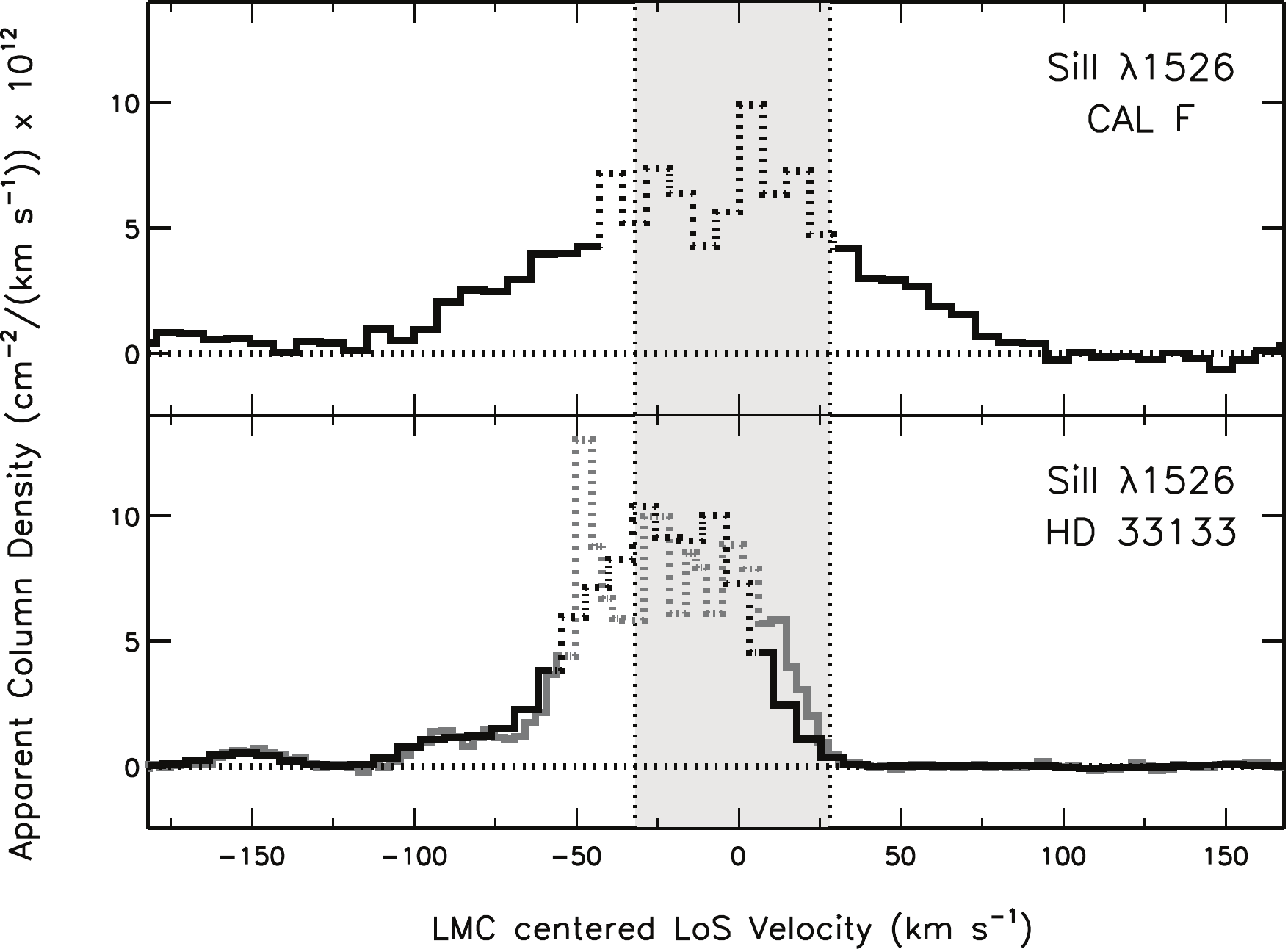} 
\includegraphics[scale=0.425,angle=0]{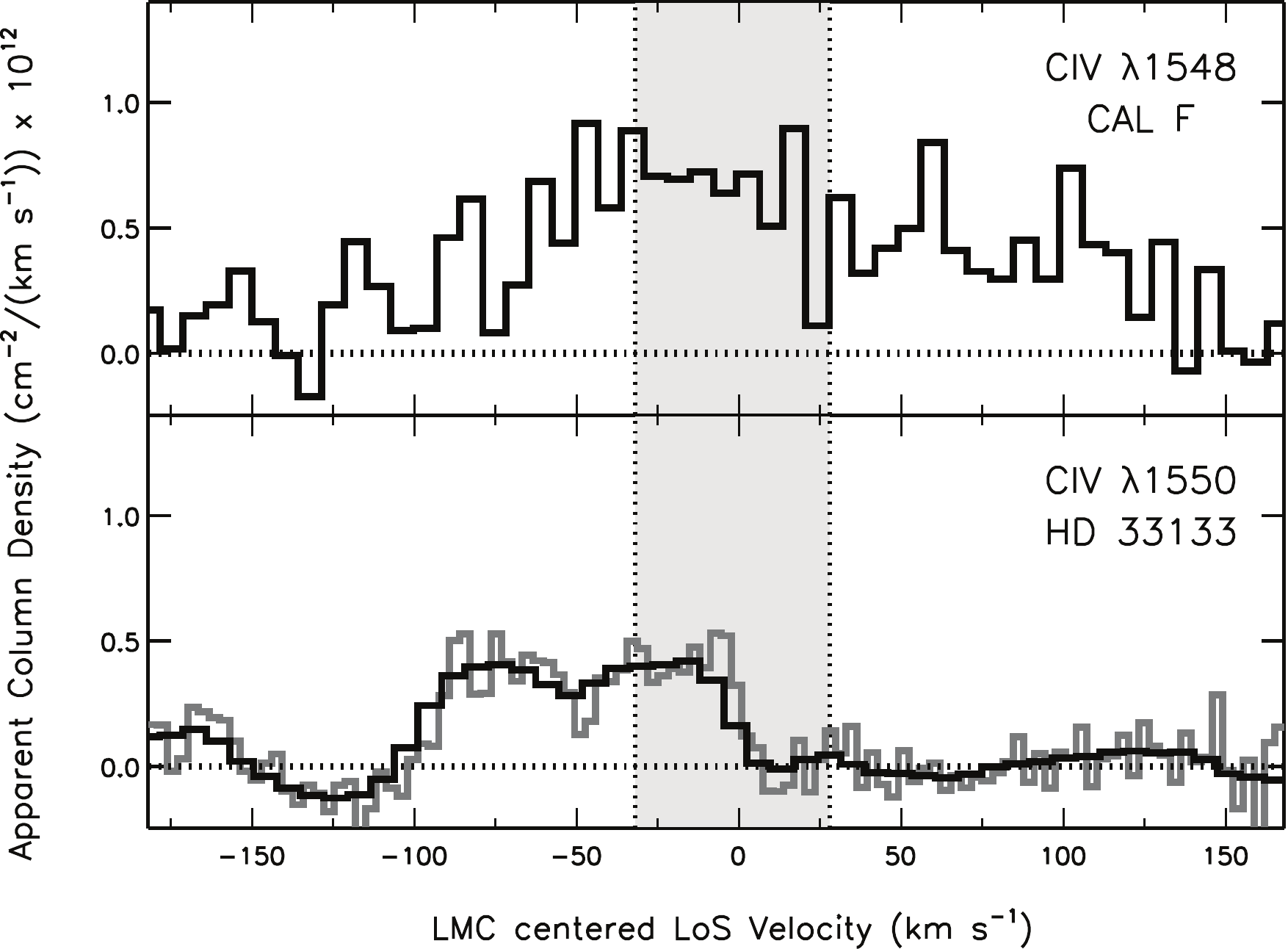} \\
\end{center}
\figcaption{Comparison of the low- and high-ionization species of the HD~33133 (right) and CAL~F (left) sight lines. The bottom panels show the full STIS $6.5~\kms$ resolution (grey) and smoothed to a $17.0~\kms$ resolution (black). The systemic velocity of the LMC at $v_{\rm LMC}=0~\kms$ is defined to be at the peak position of the H\textsc{~i} emission ($\vlsr=+282~\kms$). The extent of H\textsc{~i} emission from the LMC disk is highlighted in light grey over the $-32\le v_{\rm LMC}\le+28~\kms$ ($+250\le\vlsr\le+310~\kms$) velocity range. The O\textsc{~i} and Si\textsc{~ii} absorption saturates over $-57\lesssim v_{\rm LMC}\lesssim+18~\kms$. 
\label{figure:resolution}}
\end{figure*}

\subsection{Ionization Properties of the LMC Gas Along CAL~F}\label{section:ion_fraction}

The ionization ratios of the low- and high-ionization species along the CAL~F sight line are strikingly uniform. As Figure~\ref{figure:OI_SiII_CIV_SiIV} shows, the $N_{\rm Si\textsc{~ii}}/N_{\rm O\textsc{~i}}$ and $N_{\rm C\textsc{~iv}}/N_{\rm Si\textsc{~iv}}$ ratios are consistently $1/5$ and $3.5$, respectively, over their combined $\sim200~\kms$ velocity extent (at the \hi\ disk velocities, O\textsc{~i} and Si\textsc{~ii} are saturated). The uniformity of these ratios in the approaching and receding gas suggests the LMC's the near and far side gas flows have very similar properties and origins.  

\subsubsection{Weakly ionized gas and ionization level}

To determine the ionization properties of the LMC component, we first compare the Si\textsc{~ii} and O\textsc{~i} ions, tracers of the low ionization and neutral gas. As silicon and oxygen have similar nucleosynthetic evolution, only differences in ionization and depletion should affect the $N_{\rm Si\textsc{~ii}}/N_{\rm O\textsc{~i}}$ ratio. However, because these elements are only mildly depleted in diffuse gas, $[{\rm Si}\textsc{~ii}/{\rm O}\textsc{~i}]$\footnote{We use the familiar square-bracket notation $[{\rm X}/{\rm Y}]=\log{\left({\rm {\rm N}_X}/{\rm N}_{Y}\right)}-\log{\left({X}/{Y}\right)}_\sun$, where the solar abundance is adopted from \citet{2009ARA&A..47..481A}.} provides constraints on the hydrogen ionization fraction $(x({\rm H^+})$; \citealt{2008ApJ...679..460Z}). If $[{\rm Si}\textsc{~ii}/{\rm O}\textsc{~i}]>0$, then the gas is substantially ionized, as Si\textsc{~ii} traces both the neutral and ionized gas, whereas O\textsc{~i} only traces the neutral gas. We find $[{\rm Si}\textsc{~ii}/{\rm O}\textsc{~i}]=+0.50\pm0.08$ for the foreground and background intermediate-velocity gas surrounding the LMC; this implies an ionization fraction of $x({\rm H^+})\ge72\%$ for warm gas traced by the low-ionization species. Since highly-ionization states of silicon are not taken into account (esp. Si\textsc{~iii}), this is a strict lower limit on the total ionization fraction. 

\begin{figure}
\begin{center}
\includegraphics[scale=0.45,angle=0]{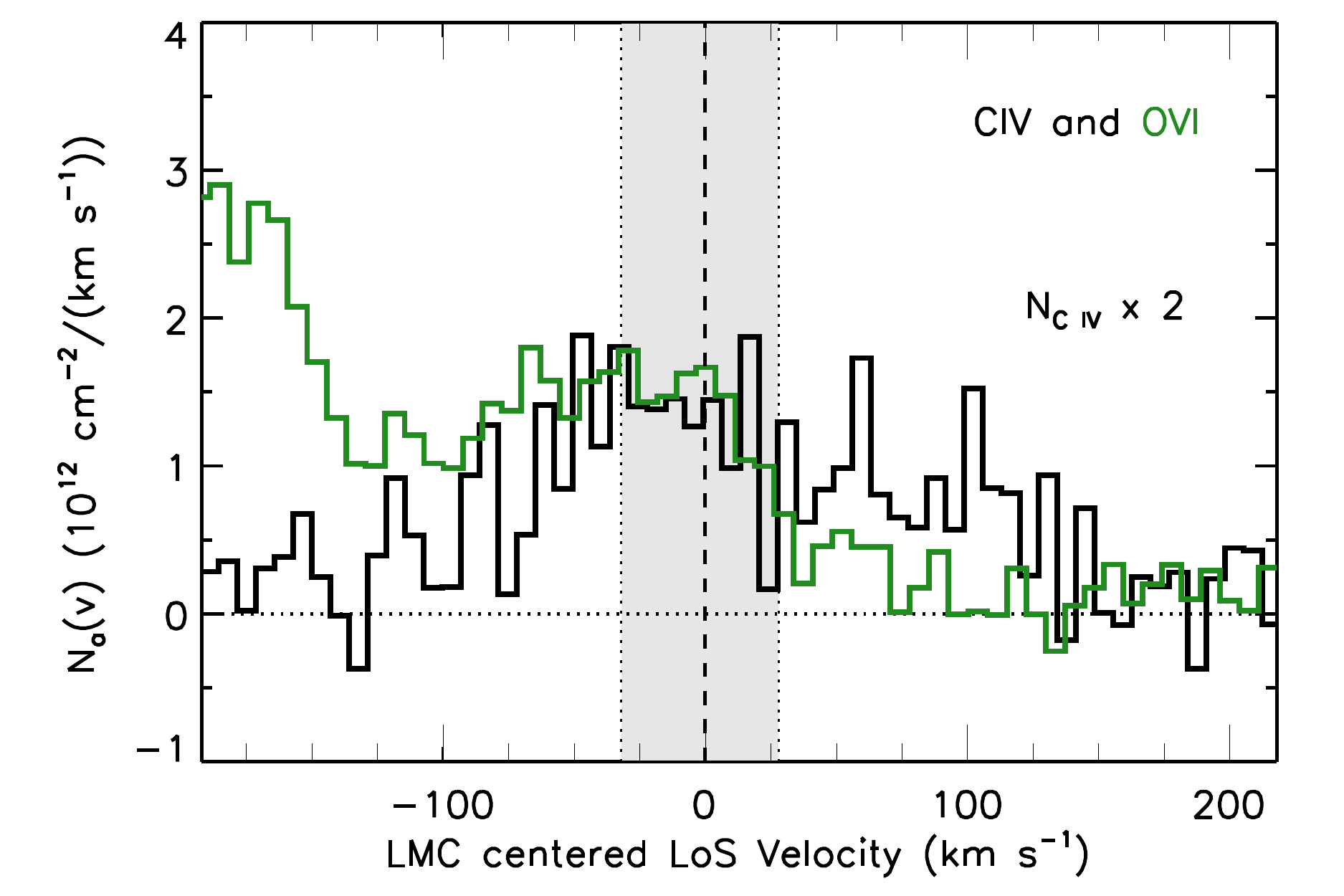}
\end{center}
\figcaption{Comparison of C~\textsc{~iv} (black) apparent column density along the CAL~F AGN sight line and of O\textsc{~vi} (green) along the stellar HD~33133 sight line. The \textit{FUSE} O~\textsc{~vi} spectrum, with a velocity resolution of $\sim15~\kms$, has been rebinned to match the C\textsc{~iv} spectrum acquired with COS a $17.0~\kms$ resolution. The O\textsc{~vi} absorption at $v_{\rm LMC}\lesssim-100~\kms$ probes the Galactic halo. The extent of H\textsc{~i} emission from the LMC disk is highlighted in light grey.
\label{figure:CIV_OIV}}
\end{figure}

\subsubsection{Highly ionized gas}

The LMC gas along the AGN CAL~F sight line is dominantly ionized with very broad absorption in the high ionization species as shown in Figure~\ref{figure:resolution} for the C\textsc{~iv} and Si\textsc{~iv} apparent column density profiles. Since low and high ions and neutral gas are observed at similar velocities, the extended LMC absorption toward CAL~F is multiphase. Unfortunately, the \textit{FUSE} observations of the CAL~F sight line have too low S/N to place any useful constraints on the O\textsc{~vi} along this sight line. There are, however, good \textit{FUSE} observations of the HD~33133 sight line. In Figure~\ref{figure:CIV_OIV}, we compare the O\textsc{~vi} toward HD~33133 and C\textsc{~iv} toward CAL~F and between $-100\lesssim v_{\rm LMC}\lesssim+25~\kms$ and find that there is an excellent match in their kinematic structure with only some minor variations, which could in part be caused by uncertainty in the O\textsc{~vi} continuum modeling (\citealt{2002ApJ...569..214H}). As the ionization potential of O\textsc{vi} is $>113~\rm{eV}$ and the absorption in the high ions (C\textsc{~iv}, Si\textsc{~iv}, and O\textsc{~vi}) is broad,  this reinforces that the gas is multiphase and also strongly suggests that the high ions are tracing a hot extended gas flow as we further argue below. The Si\textsc{~iv} and C\textsc{~iv} along the HD~33133 sight line also likely have a broad underlying component structure that is blended with narrow components that are associated with a wind nebula and the photoionized gas that surrounds this WR star as further discussed in the Appendix, which is consistent with similarly blended broad and narrow profiles that \citet{2007MNRAS.377..687L} found towards other LMC stars. As noted in \citet{2007MNRAS.377..687L} and observed toward HD~33133, there are none of the O\textsc{~vi} narrow component features that are observed in C\textsc{~iv} or Si\textsc{~iv} toward that star, implying that any O\textsc{~vi} absorption from the wind nebula is negligible.

The absence of narrow components in the C\textsc{~iv} and Si\textsc{~iv} toward CAL~F and O\textsc{~vi} toward HD~33133 and the similarity between these O\textsc{~vi}, C\textsc{~iv}, Si\textsc{~iv} profiles most certainly indicate that the high ionization species is collisionally ionized. Further insights can be gained from the ionic ratios. Over $200~\kms$ along the CAL~F sight line, the LMC C\textsc{~iv} and Si\textsc{~iv} column density profiles match each other extremely well with $N_{\rm C\textsc{~iv}}/N_{\rm Si\textsc{~iv}}\approx3.5$ (see Figure~\ref{figure:OI_SiII_CIV_SiIV}). The absence of variation in the $N_{\rm C\textsc{~iv}}/N_{\rm Si\textsc{~iv}}$ ratio strongly suggests that the same ionization processes affect the highly ionized gas over the entire velocity of the extended gaseous structures and the LMC's disk. While this is quite different from what is typically observed in the MW disk on several kiloparsec length-scale \citep{2011ApJ...727...46L}, \citet{2007MNRAS.377..687L} also noted a striking similarity between the $N_{\rm O\textsc{~vi}}/N_{\rm C\textsc{~iv}}$ ratios for the broad LMC and high-velocity components. 

Only photoionization by a hot plasma or collisional ionization process (e.g., cooling hot gas, turbulent mixing layers, and fast radiative shocks) can produce such a high C\textsc{~iv}/Si\textsc{~iv} ratio (see models by \citealt{2007ApJS..168..213G, 2009ApJ...693.1514G}, \citealt{2010ApJ...719..523K}, \citealt{2013MNRAS.434.1043O}, and observations by \citealt{2003ApJ...592..964K}, \citealt{2011ApJ...727...46L}). Further, assuming that the O\textsc{~vi} absorption toward CAL~F and HD~33133 is similar as the lie only a projected distance of $105~{\rm pc}$ appart, we have $N_{\rm O\textsc{~vi}}/N_{\rm C\textsc{~iv}}\approx2$ (see Fig. 9). This is inconsistent with photoionization by a hot plasma as this process produces ratios of $N_{\rm O\textsc{~vi}}/N_{\rm C\textsc{~iv}}\ll 1$ (see Knauth et al. 2003). Therefore the highly ionized LMC gas observed toward CAL~F is hot and collisional ionized. Since the $b$-value of C\textsc{~iv} is $\langle b\rangle = 84 \pm 9~\kms$, this would imply a temperature of $7\times10^6{\rm K}$ if the broadening was dominated by a single thermally broadened component. This is unrealistic since it would imply that the fraction of C\textsc{~iv}/C or Si\textsc{~iv}/Si is too small for the observed strong absorption of  C\textsc{~iv} or Si\textsc{~iv} (see \citealt{2013MNRAS.434.1043O}). The broadening of the C\textsc{~iv} and Si\textsc{~iv} absorption must therefore be dominated by non-thermal motions or the result of several thermally broadened ($b_{\rm th}\approx10$--$15~\kms$) components. The very symmetric high-ion kinematic profiles and absence of structures in the profiles strongly suggest that non-thermal motions is the main source of the broadening. 

We discuss below the possible origin of this broad absorption and the inferred properties of the gas along the CAL~F sight line as well as toward other sight lines throughout the LMC disk (see \citealt{2002ApJ...569..214H} and \citealt{2007MNRAS.377..687L} for these other sight lines). 
\\

\section{Discussion}\label{section:discussion}

\subsection{Origin of the Extended LMC Absorption Toward CAL~F}\label{section:cal_f_absorption}

The differences in the absorption along the CAL~F and HD~33133 sight lines  allow us to differentiate between the gas flows on the near and far side of the LMC. Although we do not know how far into the disk star HD~33133 lies, this sight line only probes foreground gaseous structures, whereas the AGN CAL~F sight line also probes the material behind the galaxy. We find that predominantly ionized gas flows away from the \hi\ disk at velocities up to $v_{\rm LMC}\approx-120~\kms$ ($\vlsr\approx+165~\kms$) on the near side of the LMC and $v_{\rm LMC}\approx+135~\kms$ ($\vlsr\approx+410~\kms$) on the far side of the LMC along the CAL~F sight line. Across the face of the galaxy, \citet{2002ApJ...569..214H} and \citet{2007MNRAS.377..687L} detected multiphase IVC gas on the near side of the LMC towards 19 sight lines. Within $100~\kms$ of the LMC's disk, the IVC absorption is relatively continuous. Multiple processes could produce these gas structures, including a lagging thick disk, tidally disrupted gas, and galaxy outflows. Here we discuss the likelihood of each of these origins for the LMC IVC gas that extends $\sim100~\kms$ off both sides of the LMC's \hi\ disk as shown in Figure~\ref{figure:resolution}. In the Appendix, we discuss the origin of the LMC HVC at $v_{\rm LMC}\lesssim-150~\kms$.

In many galaxies, a rotating thick disk gas lags behind the rotation of a thin disk (see \citealt{2008A&ARv..15..189S} and references therein). This lag has been observed as an offset of the thick disk rotation with respect to the thin disk at the same projected radial distance, sometimes including a gradient in velocity with height. This lag has been observed in both neutral gas (e.g., \citealt{2007AJ....134.1019O}) and ionized gas (e.g., \citealt{2007ApJ...663..933H} and \citealt{2007A&A...468..951K}). The kinematic differences between the thin disk, the thick disk, and any extended gas structures will cause the absorption from these structures to span different, and often larger, velocity ranges.

Any effects of differential rotation between thin disk and out of disk structures have a maximum observation signature for edge on galaxies, those with inclination angles that approach $i=90\arcdeg$. The LMC is viewed nearly face on, with an inclination angle of only $i=22\arcdeg$ (derived for the \hi\ disk gas at the position of our sight lines of interest; \citealt{1998ApJ...503..674K}). Because the line-of-sight projection of the azimuthal motions scale with $\sin{i}$, the HD~33133 and CAL~F observations are not very sensitive to the differences in azimuthal motions between its thin and thick disk. \citet{1998ApJ...503..674K} measured the maximum rotational motion of the H\textsc{~i} gas in the LMC  to be $v_{\rm rot,~max}=65~\kms$. If the vertical velocity dispersion of the ionized gas in the thick disk is roughly $10-25\%$ of the maximum rotational velocity of the LMC, then the lag along the line of sight would translate to a broadening of only $3-6~\kms$. Therefore, the spectral broadening due to a lagging thick disk is negligible.

Galaxy interactions between the MCs---and possibly the MW---have displaced over two billion solar masses of both neutral and ionized gas from the Magellanic Clouds, which now lies in the Leading Arm, Magellanic Bridge, and the Magellanic Stream (e.g., \citealt{1998Natur.394..752P, 2005A&A...432...45B, 2008ApJ...678..219L, 2009ApJ...702..940L, 2013ApJ...771..132B, Barger_MS, 2014ApJ...787..147F}). Through component fitting of kinematically resolved \hi\ observations, \citet{2008ApJ...679..432N} found that two major filamentary structures in this tidal debris could be traced back to the bridge connecting these two galaxies and the 30~Doradus starburst region of the LMC. The filament that crosses the LMC has a velocity very similar to that of the galaxy. The HD~33133 and CAL~F sight lines, however, lie far from these filamentary structures and have a much wider kinematic extent; therefore the high-velocity absorption along these sight lines is likely unassociated with the major tidal structures that are protruding from the LMC.  Further, tidal processes are unlikely to produce such symmetric velocity profiles (see Section~\ref{section:vel_collisional}) and ionization properties (see Section~\ref{section:ion_fraction}) as observed for gas on the near and far side of the LMC. 

Neither a thick disk that lags behind a thin disk nor a tidal structure that projects out of the disk can produce the absorption signatures that extend $\sim100~\kms$ beyond the \hi\ disk of the LMC. Because of the similar kinematics and ionization properties of the outflowing gas on the near side and far side of the galaxy, and because of the similarities they share with the widespread outflows found by \citet{2002ApJ...569..214H} and \citet{2007MNRAS.377..687L} on the near side of the LMC, we conclude that a large-scale wind driven by the cumulative star formation of the LMC has expanded across much of the galaxy. The lack of stellar sight lines that show signatures of inflows signifies that gas recycling through galactic fountain processes is likely inefficient (\citealt{2002ApJ...569..214H}, \citealt{2007MNRAS.377..687L}, and \citealt{2009ApJ...702..940L}), presumably because the ejected gas is more susceptible to ram-pressure stripping by the Galactic halo and tidal processes as it lies further from the center-of-mass of the LMC (see the recent works by  \citealt{2015ApJ...813..110H} and  \citealt{2015arXiv150707935S} for evidence of ram pressure effects exerted by the hot gas in the Milky Way on the Magellanic system).

\subsection{Properties of the LMC Outflows}\label{section:properties}

Along the AGN CAL~F and LMC disk star HD~33133 sight lines, we detect broad absorption that spans $\pm100~\kms$ around the \hi\ emission from the  LMC's disk (illustrated in Figure~\ref{figure:resolution}). By comparing the kinematics along these sight lines, in Section~\ref{section:vel_diff} we showed that this gas is moving away from both sides of this galaxy at speeds up to $100~\kms$ (Section~\ref{section:vel_diff}). The properties of the gas flows on both sides of the disk are also similar in the ionization fraction of the low-ionization species (Section~\ref{section:ion_fraction}) and the collisional ionization processes that are ionizing the high ionization species (Section~\ref{section:vel_collisional}). These similarities suggest that gas on the near and the far side may have the same origin. 

Much of this outflowing gas could escape from the LMC if it reaches velocities in excess of $\sim2^{1/2}$ times greater than the maximum rotational velocity of the galaxy (e.g., see \citealt{2000ApJS..129..493H}). If the outflows along the HD~33133 and CAL~F sight lines are perpendicular to the \hi\ disk, then the expelled gas velocities could exceed $\sim110~\kms$ (correcting for the line-of-sight projection), potentially enough for some of this gas to escape ($v_{\rm esc}\approx90~\kms$ with $v_{\rm rot,~max}=65~\kms$ for the H\textsc{~i}; \citealt{1998ApJ...503..674K}). Even more gas could escape when combined with tidal and ram-pressure stripping processes. These outflows could drive gas far above the galaxy's disk to create large, gaseous complexes around the LMC that feed the MW halo. 

To estimate the baryonic and metal masses and flow rates of the LMC outflows, we proceed as follows. At a given radius $(R)$, this mass is the product of the average particle mass ($\mu$), the column density perpendicular to the disk of the LMC ($N_\perp=N_{\rm obs}\cos{i}$), and the surface area ($A=\Omega f_\Omega R^2$) of the outflow, where $\Omega$ is the solid angle subtended by the wind with a covering fraction $f_\Omega$; this follows a similar procedure used by \citet{2007AA...473..791F}. The high detection rate of these outflows in active and quiescent regions of the LMC as found by \citet{2002ApJ...569..214H}, \citet{2007MNRAS.377..687L}, \citet{2011MNRAS.412.1105P}, and this study indicate that this intermediate-velocity gas covers most of the galaxy. For simplicity, we assume that $f_\Omega=1$. For comparison, \citet{2009ApJ...702..940L} found $f_\Omega=0.7$ for the low-ionization species and $f_\Omega=0.9$ for the high-ionization species of the HVC complex in the foreground of the LMC. Readily-identifiable winds observed in other galaxies typically have a biconical structure as they are often produced in the central regions of the galaxies where high concentrations of star formation and AGN activity occur (e.g., \citealt{1990ApJS...74..833H, 1993ApJ...407...83S, 1994ApJ...433...48V}). However, in the LMC, the galactic winds are likely pervasive throughout the disk as a result of its wide spread star formation; we therefore, conservatively, adopt $
\Omega=2\pi$ when calculating the mass loss due to these outflows:
\begin{equation}\label{eq:mass}
M_{\rm out}\approx\mu~N_{\rm H\perp}~2\pi~R_{\rm out}^2. 
\end{equation}
 If the ejected material travels at a constant velocity, then the rate ($\dot{M}=dM/dt$) at which the gas leaves the galaxy is given by:
 \begin{equation}\label{eq:rate}
 \dot{M}_{\rm out} \approx \mu~N_{\rm H\perp}~4\pi~R_{\rm out}~v_{\rm out}, 
\end{equation}
where $R_{\rm out}=v_{\rm out}~t$. 

To estimate the mass that has been ejected from the LMC and the corresponding rate at which it leaves the disk, we assume that the material has reached a distance equal to the \hi\ radius of LMC's disk ($R=3.7~\kpc$; \citealt{1998ApJ...503..674K}) and that it travels at an outward velocity of $v_{\rm out}=50~\kms$ (roughly half of the full extent of the continuous intermediate-velocity absorption). These assumptions translate into an outflow time period of $75~{\rm Myrs}$. 

\begin{deluxetable}{lcc}
\tabletypesize{\footnotesize}
\tablecolumns{3}
\tablecaption{Component Summary of the Ionization Species along the CAL~F sight line\label{table:columns}}
\tablehead{
 \colhead{Ion Transition} & \colhead{$\log{N_{\rm ion}/\cm^{-2}}$} & \colhead{$\log{N_{\rm ion}/\cm^{-2}}$} \\
\colhead{} & \colhead{$[-117,-32]$\tablenotemark{a}} & \colhead{$[+33, +128]$\tablenotemark{a}}
} \startdata
$\rm O\textsc{~i}~\lambda1302$   	 & $>15.13$	& $>14.84$ \\
$\rm Si\textsc{~ii}~\lambda1526$  	& $>14.47$	& $14.20\pm0.04$ \\
$\rm Si\textsc{~iii}~\lambda1206$ 	& $>13.79$	& $>13.68$ \\
$\rm Si\textsc{~iv}~\lambda1393$ 	& $13.14\pm0.04$	& $13.02\pm0.04$ \\
$\rm Si\textsc{~iv}~\lambda1402$ 	& $12.94\pm0.12$	& $13.10\pm0.09$ \\
$\rm C\textsc{~iv}~\lambda1548$ 	& $13.60\pm0.07$	& $13.78\pm0.05$ \\
$\rm C\textsc{~iv}~\lambda1550$ 	& $13.47^{+0.13}_{-0.20}$	& $13.89\pm0.08$ 
\enddata 
\tablenotetext{a}{Integration range (in $\kms$) in the LMC frame. This corresponds to $+165\le \vlsr \le+250~\kms$ and $+315\le \vlsr \le+410~\kms$.} 
\end{deluxetable}

We adopt the following assumptions to estimate the mass of the baryons and metals that the LMC is losing into its surroundings and their corresponding rates: (1) An ionization fraction of $72\%$ for the gas traced by the low-ionization species; in Section~\ref{section:ion_fraction}, we found that $x({\rm H^+})\ge0.72$ based on comparisons of the O\textsc{~i} and Si\textsc{~ii} ions. (2) An ionization fraction of $100\%$ for the gas traced by the high-ionization species Si\textsc{~iv} and C\textsc{~iv}. (3) An average particle mass of $\mu\approx1.3 m_{\rm H}$ to account for the contribution of He when determining the total baryon mass in a given phase. And (4) a metallicity equal to that of the present-day LMC ($Z=0.5~Z_\sun$; \citealt{1992ApJ...384..508R}); as this material is likely associated with feedback from active-stellar regions, this metallicity is a lower limit (but it is unlikely to be much larger than a factor 2-3). We integrate the absorption profiles $-117\le v_{\rm LMC}\le-32~\kms$ (near side) and $+33\le v_{\rm LMC}\le+128~\kms$ (far side) to estimate the column densities of this wind along the CAL~F sight line and list them in Table~\ref{table:columns} for select low- and high-ionization species. 

We use the Si\textsc{~ii} and Si\textsc{~iii} ions to determine the mass of the gas flowing from LMC in the neutral and low-ionization phases. Nearly all of the silicon exists in these low-ionization states (i.e., $\left(N_{\rm Si\textsc{~ii}}+N_{\rm Si\textsc{~iii}}\right)/N_{\rm Si}\approx1$). These ions have an average column density (over the near and far side of the LMC) of $\log{(\langle N_{\rm Si\textsc{~ii}}\rangle+\langle N_{\rm Si\textsc{~iii}}\rangle})>14.42$ (see Table~\ref{table:columns}), where we integrated Equation~{(\ref{eq:Na})} over the velocity interval defined in Table~\ref{table:columns} to estimate $N_{\rm obs}$. We use the $\langle N_{\rm H}\rangle=\sum\langle N_{\rm Si}\rangle~{\left(Z/Z_{\sun}\right)^{-1}~\left(\rm Si/H\right)_{\sun}^{-1}}$ to convert these column densities into total hydrogen column density ($\log(N_{\rm H_{total}})\ge19.22$) and Equations~(\ref{eq:mass}) and~(\ref{eq:rate}) to calculate the corresponding outflow masses and outflow rates. The mass loss contribution for the low-ionization phases is given by:

\begin{equation}
M_{\rm gas,~low}=1.3 m_{\rm p}\frac{\left(\langle N_{\rm Si\textsc{~ii}}\rangle+\langle N_{\rm Si\textsc{~iii}}\rangle\right)~\cos{i}}{Z/Z_{\sun}~\left(\rm Si/H\right)_{\sun}}~2\pi R^2
\end{equation}

To determine the mass contribution of the high-ionization species ($M_{\rm gas,~high}$), we use Si\textsc{~iv}. The broad component structure of the Si\textsc{~iv} line and the $N_{\rm C\textsc{~iv}}/N_{\rm Si\textsc{~iv}}\approx3.5$ along the CAL~F sight line indicate that Si\textsc{~iv} probes collisionally ionized gas (see Sections~\ref{section:vel_collisional} and~\ref{section:ion_fraction}). The average column density of this ion along both sides of the LMC disk is $\log{(\langle N_{\rm Si\textsc{~iv}}\rangle})=13.05\pm0.03$ (see Table~\ref{table:columns}). However, because silicon will also exist in higher-ionization states, this calculation must also convert the fraction of silicon that exists in the Si\textsc{~iv} state to the total silicon in all high-ionization states (i.e., $N_{\rm Si,~{\rm high}}=\langle N_{\rm Si\textsc{~iv}}\rangle\left(N_{\rm Si\textsc{~iv}}/N_{\rm Si}\right)^{-1}$). Through photoionization and collisional ionization models, \citet{2013MNRAS.434.1043O} found a maximum Si\textsc{~iv} ionization fraction of $N_{\rm Si\textsc{~iv}}/N_{\rm Si}=0.4$---although this value could be a factor 6 lower for a wide range of typical temperatures. We adopt this maximum ionization fraction when calculating $N_{\rm Si}$, which places a strict lower limit on the total mass loss when converted to $N_{\rm H}$. 

Galactic winds that are driven by stellar activity that expels metal enriched gas into and out of the galaxy. To similarly convert from the silicon column densities to metals, we assume that $\left(m_{\rm Si}/m_{\rm Metals}\right)_\sun=0.064$ for solar mass fraction of metals in silicon \citep{2009ARA&A..47..481A}, where $\mu_{\rm Si}=28~m_{\rm H}$ (see \citealt{2014ApJ...786...54P} and \citealt{2015ApJ...804...79L} for more details on this procedure): 
\begin{equation}
M_{Z,~{\rm total}}=28 m_{\rm p}\frac{\left(\langle N_{\rm Si\textsc{~ii}}\rangle+\langle N_{\rm Si\textsc{~iii}}\rangle+\langle N_{\rm Si\textsc{~iv}}\rangle\right)~\cos{i}}{\left(m_{\rm Si}/m_{\rm Metals}\right)_\sun}~2\pi R^2 
\end{equation}
These mass losses and their corresponding rates are listed in Table~\ref{table:mass_rate}.

\begin{deluxetable}{cccc}
\tabletypesize{\footnotesize}
\tablecolumns{3}
\tablecaption{Summary of Outflow Masses and Rates\tablenotemark{a}\label{table:mass_rate}}
\tablehead{
\colhead{} & \colhead{Gas Phase} & \colhead{$M$} & \colhead{$\dot{M}$} \\
 \colhead{} & \colhead{} & \colhead{($10^6~M_\sun$)} & \colhead{$M_\sun/\yr^{-1}$}
} \startdata
Baryons	& Low~Ions 		& $\gtrsim14.6$\tablenotemark{b} 	& $\gtrsim0.41$ \\
Baryons	& High~Ions\tablenotemark{c} 		& $>1.4$\tablenotemark{e}	& $>4.0\times10^{-2}$\tablenotemark{d} \\
Metals	& Low- \& High-Ions	& $>8.0\times10^{-2}$	& $>2.2\times10^{-3}$
\enddata 
\tablenotetext{a}{We assume that these winds reach a distance of $3.7~\kpc$ and are moving at a rate of $50~\kms$. We also assume a present-day LMC metallicity of $Z=0.5~Z_\sun$ (\citealt{1992ApJ...384..508R}); if these outflows are associated with stellar feedback, then they could be enriched by as much as $2-3$ times more metals. These values are therefore conservative estimates.} 
\tablenotetext{b}{Gas probed by Si\textsc{~ii} and Si\textsc{~iii}. Both of these ions are saturated near LMC disk velocities.}
\tablenotetext{c}{For the photoionized and collisionally ionized gas probed by Si\textsc{~iv}. } 
\tablenotetext{d}{If we use C\textsc{~iv} instead of Si\textsc{~iv} to calculate $M$ and $\dot{M}$ in the highly ionized phase, where $N_{\rm C\textsc{~iv}}/N_{\rm C}$ is assumed to be $<0.3$ (the maximum expected value for collisionally and photoionized gas; see \citealt{2007ApJS..168..213G}) and ${\rm C/H}=10^{-4}$ \citep{1998RMxAC...7..202K}, we find $0.7\times10^6~M_\sun$ and $0.02~M_\sun/\yr^{-1}$.
}
\end{deluxetable}
 
We find that the galactic outflows of the LMC have ejected at least $1.6\times10^7~{\rm M}_{\sun}(R_{\rm out}/3.7~\kpc)^2$ of gas from its disk. For comparison, \citet{2014ApJ...787..147F} found that tidal interactions have removed more than $2.0\times10^9~{\rm M}_{\sun}$ in neutral and ionized gas from both of the Magellanic Clouds. Although we find that the mass of the LMC's outflow is $100$ times less than that in tidal debris surrounding these two galaxies, we are only viewing this distribution for one snapshot in time. The galactic winds from the LMC could have ejected more than $10\%$ of the total mass and more than $60\%$ of the total metals found in these tidal structures if they have been relatively continuous since the Magellanic Stream was created $1-2~{\rm Gyr}$ ago (e.g., \citealt{2010ApJ...721L..97B}). As the global star formation activity in the LMC has not been continuous, the galactic winds likely died down when the stellar activity diminished. The global star formation of the LMC peaked at $12~{\rm Myr}$, $100~{\rm Myr}$, $500~{\rm Myr}$, and $2~{\rm Gyr}$ with the last two bursts coinciding with a burst star formation in the SMC, suggesting that they were induced by interactions between the two galaxies \citep{2009AJ....138.1243H}. Surrounding the Magellanic Clouds, we see evidence for multiple outflow episodes with the LMC HVC complex at $\vlsr=+130~\kms$ (see \citealt{2009ApJ...702..940L}) and an enriched Magellanic Stream filament (see \citealt{2013ApJ...772..111R}). 

Galactic winds generated by stellar activity have been observed in other galaxies. \citet{1998ApJ...506..222M} detected outflows in 13 out of 15 of their star-forming dwarf galaxies with masses and specific star-formation rates (sSFR) ranging from $8.1\le \log(M_{\star}/M_{\sun})\le10.4$ and $5\times10^{-12}\le {\rm sSFR} \le8\times10^{-10}~\yr^{-1}$ (these ranges exclude M82 as it is well-known for having exceptionally strong outflows). By summing together all of the ionized gas mass detected in \ha\ shells expanding out of the galaxies' disks, they find wind masses of $8\times10^{3}\le M_{\rm out}/M_{\sun}\le3\times10^{6}$ that are flowing at a rate of $2\times10^{-3}\le(\dot{M}_{\rm out}/M_{\sun}\yr^{-1}) \le2$; these values assume a volume filling factor of $\epsilon=0.01$, which is consistent for pressure equilibrium conditions between the warm and hot gas in these shells and is expected to be accurate to within a factor of $10$ \citep{1999ApJ...513..156M}. The span of their galaxy sample includes galaxies with LMC masses and SFR, where the LMC has ${M}_{\rm \star}\approx3\times10^{9}~{M_\sun}$ \citep{2009IAUS..256...81V} and  ${\rm SFR}_{\rm LMC}\lesssim0.2~M_{\sun}~\yr^{-1}$ (or ${\rm sSFR}=2\times10^{-11}~\yr^{-1}$) (\citealt{2009AJ....138.1243H}). Their outflow masses and rates are consistent with the values we estimate for the LMC of $M_{\rm out}=1.3\times10^{6}M_{\sun}$ and $\dot{M}_{\rm out}=4\times10^{-2}~M_{\sun}\yr^{-1}$, assuming $R_{\rm out}=3.7~\kpc$ and $v_{\rm out}=50~\kms$. 
 
Both of the Magellanic Clouds have lost substantial quantities of metals over their lifetimes. Similarly, substantial metals have been detected around other nearby dwarf galaxies (\citealt{2014ApJ...796..136B} and \citealt{2014MNRAS.445.2061L}). Within a distance of one half the virial radius ($15~\kpc$ to $100~\kpc$ for galaxy masses $8.5\le\log{M_\star/M_{\sun}} \le10$) and within $150~\kms$ of the systemic velocity of their isolated dwarf galaxy sample, \citet{2014ApJ...796..136B} found that there is more carbon surrounding these dwarf galaxies than locked in their stellar populations, with carbon masses that exceed $10^6~{\rm M}_\sun$.  The stark difference in carbon mass between their galaxy sample and the LMC at only few hundred solar masses could be related to the difference in the volume probed where their C\textsc{~iv} detections span out to $\sim100~\kpc$ from their galaxies whereas we are likely probing only a few $\kpc$ to tens of $\kpc$ for the gas within $\sim100~\kms$ of the LMC (the LMC is $50~\kpc$ away). Extending $R_{\rm out}$ to $100~\kpc$ and assuming that $f_\Omega=1$ and $\Omega=2\pi$, the total carbon mass around the LMC would still only be $\sim1/20$ of that found by the \citet{2014ApJ...796..136B}. However, their observations are tracing the amount carbon that surrounds their galaxies sample regardless of how that material got there and therefore traces multiple pollution processes in addition to outflows. Additionally, the environments of these galaxies are very different---the Magellanic Clouds have lost over $2\times10^9~{\rm M}_\sun$ due to galaxy interactions \citep{2014ApJ...787..147F}, while the \citet{2014ApJ...796..136B} galaxies are isolated. Furthermore, any outflows may be further stripped away from the LMC by interactions with the MW corona.

The regions surrounding galaxies have been observed to be a reservoir of gas and metals  that is strongly correlated with the star-formation activity within them (e.g., \citealt{2011Sci...334..948T} and \citealt{ 2014ApJ...796..136B, 2014ApJ...794..130B}). In $L^*$ galaxies, roughly half of the metals \citep{2014ApJ...786...54P} and baryons \citep{2014ApJ...792....8W} exists outside of their disks. \citet{2011ApJ...743...10B} and \citet{2014ApJ...794..156R} find that the detection rate of the galactic winds is strongly correlated with the inclination angle of the galaxy, where this rate is greatest for near face-on galaxies for a sample of 105 galaxies that spans a mass range of $9.5\lesssim \log{M_{\star}/M_{\sun}}\lesssim11.5$ at $z>0.3$. The LMC is both near face-on and forming stars, but has an outflow rate for the low-ionization species that is more than a magnitude lower than the rate found for the star-forming galaxies in \citet{2014ApJ...794..156R}. However, the LMC is on the low-mass end of their sample and a lower-mass sample of galaxies is needed for a direct comparison with the LMC outflow quantities. \\

\subsection{The HVC complex toward and around the LMC}\label{section:HVC}

We noted in Section~\ref{section:vel_collisional} that there is detection of a high-velocity absorber at $v_{\rm LSR}\approx +130~\kms$ toward CAL~F and  HD~33133 (see Figures~\ref{figure:HD33133_norm} and~\ref{figure:cal_f_norm}). The HVC absorption is less prominent along the CAL~F sight line as the COS spectrograph has a lower velocity resolution than the STIS spectrograph (see Table~\ref{table:observations}), but it is present in the ${\rm C}\textsc{~ii}~\lambda1334$, ${\rm Si}\textsc{~ii}~\lambda1260,~1526$, and ${\rm Si}\textsc{~iii}~\lambda1206$ transitions. For this cloud, we measure $\left[{\rm Si}\textsc{~ii}/{\rm O}\textsc{~i} \right]=+0.25\pm0.12$, placing its ionization fraction of the low-ionization gas of $>55\%$. This HVC is part of the well studied HVC complex toward the LMC at  $+90\la \vlsr \la +175 $ \kms\ that spans at least the angular extent of the LMC with a $70\%$ covering fraction in ${\rm O}\textsc{~i}~\lambda1039$, ${\rm Fe}\textsc{~ii}~\lambda1144$, and ${\rm N}\textsc{~ii}~\lambda1083$ and 90\% in ${\rm O}\textsc{~vi}~\lambda1031$ \citep{2009ApJ...702..940L}. It has an ionization fraction that exceeds $50\%$ for the low-ionization species \citep{2009ApJ...702..940L}, consistent with the ionization level observed toward CAL~F and HD~33133. 

The HVC at $+90\lesssim\vlsr\lesssim+175~\kms$ has been detected in both \hi\ emission \citep{2003MNRAS.339...87S} and UV absorption \citep{2007MNRAS.377..687L,2009ApJ...702..940L}. Several properties of this HVC connected its origin to an outflow from the LMC: (1) \citet{2003MNRAS.339...87S} found that the regions of this complex with high \hi\ column densities are often projected onto \hi\ voids (such as supergiant shells, e.g., LMC\,3) within the LMC disk that connect back to the disk of the LMC with spatial and kinematics bridges (see their Figures~9 and~11). \citet{2003MNRAS.344..741R} also found similar bridges in $\left[{\rm O}\textsc{~iii}\right]$ emission towards the 30~Doradus starburst region (see their Figures~4--6). (2) This complex shows a gradient in its LMC-frame velocity with RA across the entire LMC in both H\textsc{~i} emission as well as O\textsc{~i} and Fe\textsc{~ii} absorption (see Figure~5 in \citealt{2009ApJ...702..940L}). (3) This HVC complex has a metallicity that is similar to that of the LMC ($\left[{\rm O}/{\rm H}\right]=-0.4$; e.g., \citealt{1992ApJ...384..508R}) with $\left[{\rm O}/ {\rm H}\right]=-0.51^{+0.12}_{-0.16}$ \citep{2009ApJ...702..940L}. And (4) it contains dust based on the depletion of iron relative to silicon (\citealt{2009ApJ...702..940L}, and see also \citealt{1999ApJ...512..636W, 2015MNRAS.451.4346S}) in addition to H$_2$ \citep{1999Natur.402..386R}, although molecules are not widespread (see \citealt{2009ApJ...702..940L}). 

However, HVC absorption at $v_{\rm LSR}\approx +150~\kms$ in this general direction has now been detected toward a star for which the distance is estimated to be $d=9.2\,^{+4.1}_{-7.2}$ kpc from the Sun \citep{2015arXiv150908942W}. This provides an upper limit on the distance to the HVC complex toward the LMC of $<13$ kpc. Clearly, a present-day outflow from the LMC is not a plausible origin anymore in the light of this distance as argued by \citet{2015arXiv150908946R}. However, this HVC complex is  quite peculiar compared to the overall HVC population with the presence of dust (uncommon in HVCs; see \citealt{1986A&A...170...84W} and \citealt{2001ApJS..136..463W}). If this HVC complex is not related to the LMC, the kinematic and morphological properties described above that appear to connect it to the LMC would be entirely fortuitous. 

It is quite plausible that this complex is just a Milky Way HVC. However, given its peculiarities compared with most HVCs, we explore whether this complex could represent an ancient LMC outflow driven by a burst of star formation some time ago. Assuming an average velocity of 100--150 \kms, it would take to this gas about 250--400 Myr to travel $40~{\rm kpc}$ (ignoring possible drag forces and initial higher velocity). Coincidentally, this travel time corresponds to an epoch when a burst of star formation occurred in the LMC \citep{2009AJ....138.1243H}. This is quite far for an HVC to travel through the ionizing radiation fields of the surrounding galaxies and through the Milky Way's hot halo. The survival of an HVC is dependent on its size, mass, its initial velocity, as well as if it is shielded by its magnetic field. \citet{2009ApJ...698.1485H} show that for  typical cloud velocities and halo densities, H\textsc{~i} HVCs with masses $<10^{4.5}~{\rm M}_{\sun}$ will lose their H\textsc{~i} content within $10~{\rm kpc}$ or less; however, their remnants could contribute warm ionized HVCs. Armillotta, Fraternali, and Werk (2015, in prep.) are investigating the survival of an HVC with LMC metallicity traveling through the hot MW halo with different initial masses, sizes, and velocities, and early results suggest that massive clouds of $>10^6~{\rm M}_\sun$ are far more difficult to disrupt than smaller cloud, i.e., they may be able to survive for several hundreds of Myrs. This is because in the low density gas of the MW halo, both the ram pressure and Kelvin-Helmholtz instability effects are weak, which considerably slows down cloud disruption (see also \citealt{2015MNRAS.447L..70F}). 

If the LMC had a major outflow some 300--400~${\rm Myr}$ ago, we would expect that (1) the HVC complex has expanded beyond the disk of the LMC and (2) toward QSOs a higher velocity counterpart corresponding to the receding ancient outflow might be present. While it is beyond the scope of this paper to do a full analysis of the  other QSOs in our HST program 11692 and other QSOs since observed with other programs, thanks to the survey by \citet{2014ApJ...787..147F}, all these QSO spectra are available. In Table~\ref{table:hvc}, we list QSOs within $20\degr$ of the LMC's kinematic center\footnote{$(l,b) = (279.75\arcdeg, -33.60\arcdeg)$; Kim et al. (1998).} showing high-velocity absorption components (excluding QSOs toward the Magellanic Bridge). In the Fox et al. paper, these QSOs are identified as LMC-On and LMC-Off. As the \hi\ LMC disk radius is about $4.5\degr$, this sample of QSOs allows us to search for high-velocity absorption components well beyond the LMC disk,  $2.8\degr$ to $19.3\degr$ from the LMC center. 

To identify absorption at $v_{\rm LSR} > +90~\kms$, we require that there is absorption at the same velocity for at least two different ions from the following list: Si\,{\sc ii}, Si\,{\sc iii}, Si\,{\sc iv}, C\,{\sc ii}, and/or C\,{\sc iv}. Within $17\degr$ from the LMC center, high-velocity absorption is detected at velocities $+100 \le v_{\rm LSR} \le +160~\kms$, consistent with the HVC complex velocities observed toward the LMC.\footnote{We note that the anti-correlation between the LMC standard of rest velocity and the R.A. derived by \citealt{2009ApJ...702..940L} also applies to the QSO sight lines at smaller and larger R.A., although with a larger scatter.} As noted by \citet{2014ApJ...787..147F}, all the QSO sight lines have evidence for LMC halo gas well beyond the disk of the LMC up to $20~{\rm kpc}$ at velocities $+260\lesssim v_{\rm LSR} \lesssim+315~\kms$. There is also evidence in 4/11 sight lines of very high-velocity absorption at $+360\lesssim v_{\rm LSR} \lesssim +420~\kms$. These velocities are quite peculiar, offset by 80--140~$\kms$ from a systemic velocity of $\vlsr=+280~\kms$ \citep{2003MNRAS.339...87S,1998ApJ...503..674K}. This offset relative to the LMC systemic velocity is quite similar to the HVC complex velocity difference with the LMC systemic velocity, suggesting that the very high-velocity absorption at $+360 \le v_{\rm LSR} \le +420~\kms$ could be the receding component of the ancient LMC outflow while the HVC complex at $+100 \le v_{\rm LSR} \le +160~\kms$ could be the approaching component. 

The two expected outcomes from an ancient LMC outflow are observed, providing some additional support to the scenario that an ancient LMC outflow may feed the MW lower halo. Although the properties of this high-velocity gas observed towards the LMC have many attributes that are uncommon amongst the Milky Way's cloud population, there is the possibility that it is unrelated to an LMC outflow. However, while the covering fraction of the MW HVCs seen in UV absorption with $90 \la |v_{\rm LSR}| \la 170~\kms$ is quite high, around 60\%, the covering fraction of HVC with $|v_{\rm LSR}|>170~\kms$ is only about 10\% when the Magellanic Stream is removed \citep{2012MNRAS.424.2896L}; and at these positions, this is unlikely to be Magellanic Stream gas. Future simulations of the travel of high-velocity gas through the MW halo will be able to test the survivability of an HVC complex depending on its initial mass and test if outflows from dwarf galaxies may be one of the sources of the HVC gas seen at 5--20~${\rm kpc}$ from the Sun. 

\begin{deluxetable*}{lccccl}
\tabletypesize{\footnotesize}
\tabcolsep=1pt
\tablewidth{0pc}
\tablecaption{Velocity Components with $v_{\rm LSR}\ge + 90$ \kms\ toward QSOs around the LMC \label{table:hvc}}
\tablehead{
 \colhead{QSO} & \colhead{$g_l$} & \colhead{$g_b$} & \colhead{$\Delta \theta$\tablenotemark{a}} & \colhead{Direction\tablenotemark{a}} & \colhead{$v_{\rm LSR}$}   \\
  \colhead{}   & \colhead{($\degr$)} & \colhead{($\degr$)} & \colhead{($\degr$)}  & \colhead{} & \colhead{(\kms)} 
} 
\startdata
CAL~F		   &   277.2 & $-35.4 $& 2.8&   N    & $+130, +260, +410$		\\
PKS\,0552--6402    &   273.5 & $-30.6 $& 6.1&   NW   & $+100, +280, +420$		\\
2E\,0622.9-6434	   &   274.3 & $-27.3 $& 7.8&   NW   & $+110, +280, +380, +420$	\\
RBS\,563 	   &   272.3 & $-39.2 $& 8.3&   N    & $+100, +150, +220, +300$  	\\
PKS\,0637--752 	   &   286.4 & $-27.2 $& 8.6&   SE   & $+110, +240, +320$   	\\
RBS\,542 	   &   267.0 & $-42.0 $&13.1&   NW   & $+100, +160, +310, +360$	\\
HE\,0429--5343 	   &   262.1 & $-42.2 $&16.3&   NW   & $+100, +280$		\\
RBS\,567 	   &   261.2 & $-40.9 $&16.4&   NW   & $+100, +220, +280, +315$	\\
HE0\,435--5304 	   &   261.0 & $-41.4 $&16.7&   NW   & $+100, +280$		\\
HE0\,439--5254 	   &   260.7 & $-40.9 $&16.8&   NW   & $+100, +280$		\\
PKS\,0558--504 	   &   258.0 & $-28.6 $&19.3&   NE   & $+265$		 
\enddata 
\tablenotetext{a}{Angular separation and approximate direction from the LMC center.} 
\end{deluxetable*}

In all the above, we have ignored any shear the Galactic halo gas may impart as the LMC and its outflow move through it. The LMC is moving at $\sim300~\kms$ relative to the MW \citep{2013ApJ...764..161K}. If the drag from the halo imparts even a $10~\kms$ transverse motion of the outflow relative to the LMC, it will displace it by $\sim4~\kpc$, or an LMC radius, in $400~{\rm Myr}$. Our earlier suggestion that a lack of inflowing gas may be the result of incorporation of that material into the Galactic corona relies on a similar effect. Thus, it is not clear how an outflow from several hundred Myr ago would maintain its position between us and the LMC.  Thus despite being among the best characterized HVC complex with an unprecedented level of information regarding its kinematics, metallicity, ionization and dust structures on both small and large scales, its origin remains uncertain. To make progress we see two ways. From the theoretical side, future hydrodynamical  simulations taking into account the relative motion of the LMC and MW may help providing additional clues regarding the origin of this HVC complex. From the observational side, we have started to collect Wisconsin \ha\ Mapper (WHAM) observations to map in \ha\ the true spatial and kinematical distributions of the HVC complex. This \ha\ map will inform us on its overall displacement relative relative to the LMC and geometry, and hence  we will be able to reassess its origin and estimate a more accurate mass. Preliminary analysis reveals that this HVC could extend more than 10-degrees from the center of the LMC in the direction of the RBS~542 sight line at $\vlsr\approx+175~\kms$ and more than 20-degrees in the direction of the PKS~0558-504 sight line at $\vlsr\approx+250~\kms$, consistent with the absorption listed in Table~\ref{table:hvc}. More analysis will need to be done to determine if the emission in these directions trace the same HVC complex.

We finally note that the mass of this HVC complex toward the LMC can be assessed with its newly measured distance and with a new determination of its sky coverage. This HVC complex is possibly one of the best characterized HVCs thanks to the multitude of background LMC stars used to determine its properties as described above. Following \citet{2009ApJ...702..940L}, the mass of the neutral gas in the outflow can be estimated with $M^{\rm neut}_{\rm HVC} =  1.3 m_{\rm p} f_{\rm cov}\ \overline{N}_{\rm O\textsc{~i}}\ ({\rm O/H})^{-1}_{\rm HVC}\ A_{\rm HVC}$, where the covering factor is $f_{\rm cov} = 0.7$, the average O\textsc{~i} column density is $\overline{N}_{\rm O\textsc{~i}} \simeq 10^{14.9}~\cm^{-2}$ \citep{2009ApJ...702..940L}. For the HVC complex seen only over the LMC disk, but now assuming a distance of $5~{\rm kpc}$ from the MW with an area of the HVC is $A_{\rm HVC}\simeq 1.9$ kpc$^2$, results in a new estimate of $M^{\rm neut}_{\rm HVC} \approx 7\times 10^4~{\rm M}_\sun$. Here we assume that the HVC spans the surface of a sphere for the area, but note that the geometry is very uncertain and that other reasonable geometries (i.e., rectangular, disk, etc.) could decrease this mass estimate by a factor of $3$ (see e.g., values in \citealt{2015arXiv150908946R}). Including the photoionized gas probed by Si\,{\sc ii} and Fe\,{\sc ii}, the mass would increase by at least another factor $2$, and including the O\textsc{~vi} would increase the mass by another factor $\ga 2$ \citep[see][]{2009ApJ...702..940L}, yielding a total mass of $\ga  3 \times 10^5~{\rm M}_\sun$. As discussed above, the HVC complex extends beyond the HVC \hi\ disk by an additional $\sim$10\degr\ (i.e., $\theta_{\rm dist,~HVC} \simeq 30\degr$), implying a total mass of $\ga 3 \times10^6~{\rm M}_\sun$ assuming that the properties (ionization, metallicity) of the HVC beyond the LMC disk are the same. If the HVC complex toward the LMC is an ancient LMC outflow, and if the initial mass of that ancient outflow was similar to the mass of the present-day mass outflow (see Section~\ref{section:properties}), this would mean that most of the mass of the ancient outflow has been lost before reaching the inner MW halo. As described above, ongoing and future WHAM observations will allow us to better define the geometry and extent of this HVC complex, and hence to more accurately determine its mass. 

 \section{Summary}\label{section:summary}
  
We have investigated the gas surrounding the LMC by comparing the absorption along a star embedded within its disk (target HD~33133) with a background AGN (target CAL~F). These targets are only $7.2\arcmin$ apart on the sky (corresponding to a projected physical separation of $105~{\pc}$), which enables us to probe similar large-scale environment along these sight lines and to separate the gas flows on the near and far side of the galaxy's disk. Using UV spectroscopic {\it HST} observations, we show that gaseous material is flowing away from both sides of the \hi\ disk of the LMC in Section~\ref{section:properties}. From the strikingly similar properties of the near  and far side gaseous outflows along the neighboring HD~33133 and CAL~F sight lines, the detection of similar intermediate-velocity absorption with high ${N}_{{\rm C\textsc{~iv}}}/{N}_{{\rm Si\textsc{~iv}}}$ ratio toward stars spread across the LMC in \citet{2007MNRAS.377..687L}, and the scarcity of inflowing material observed in \citet{2002ApJ...569..214H} toward stars also scattered throughout the disk, we conclude that a pervasive wind expels a large amount of predominantly ionized gas from the disk of the LMC. The outflowing material within $\sim$$100~\kms$ of the LMC has the following properties:    

 \begin{enumerate}
\item{\bf Kinematics:} We observe a kinematically continuous, multiphase gas that is made up of neutral gas and low- and high-ionization species that extends further than the kinematic extent of the \hi\ disk of the LMC. This material is flowing away from the LMC's disk with speeds up to $-100~\kms$ on the near side and $+100~\kms$ on the far side of the galaxy. The absorption is very broad, having breadths large enough to require significant non-thermal motions (see Figures~\ref{figure:resolution} and~\ref{figure:OI_SiII_CIV_SiIV}). 
\item{\bf Ionization Fraction:} The warm intermediate-velocity gas ($v_{\rm LMC}=\pm100~\kms$) on both sides of the LMC has an ionization fraction that exceeds $72\%$ ($[{\rm Si}\textsc{~ii}/{\rm O}\textsc{~i}]=+0.50\pm0.08$; see Section~\ref{section:ion_fraction}). The ionization fraction of the highly-ionized gas at greater temperatures likely approaches unity. 
\item{\bf Ionization Sources:} The gas is multiphase with the low ions being most certainly photoionized. The high ions (Si\textsc{~iv} and C\textsc{~iv}) have been predominantly ionized through collisional processes as evidenced by their broad and smooth velocity structure and an ionic ratio of $N_{\rm C\textsc{~iv}}/N_{\rm Si\textsc{~iv}}\approx3.5$ (see Section~\ref{section:ion_fraction}). 
\item{\bf Baryonic Mass and Mass-Loss Rate:} We estimate the mass and mass-loss rates of the intermediate-velocity gas within  $100~\kms$ of the LMC at $1.5\times10^7~{{\rm M}_\sun}\, (R_{\rm out}/3.7~\kpc)^2$ and $4.1\times10^{-1}~{\rm M}_\sun \,\yr^{-1} (v_{\rm out}/50~\kms)$ in the low-ionization state and $>1.4\times10^6~{{\rm M}_\sun}\, (R_{\rm out}/3.7~\kpc)^2$ and $>4.0\times10^{-2}~{\rm M}_\sun\,\yr^{-1}\, (v_{\rm out}/50~\kms)$ in high-ionization state. The low ions dominate the mass contribution for the gas-phases explored in this study. 
\item{\bf Metal Mass and Mass-Loss Rate:} We found that the LMC galactic wind is ejecting more than $8.0\times10^4~{\rm M}_\sun\, (R_{\rm out}/3.7\, \kpc)^2$ from the LMC disk at a rate exceeding $2.2\times10^{-3}~{\rm M}_\sun\, \yr^{-1}\, (R_{\rm out}/3.7~\kpc)\, (v_{\rm out}/50~\kms)$ in the low- and high-ionization states. These value will increase if these outflows contain metal enriched gas compared to the present-day metallicity of the LMC. 
\end{enumerate}

Because the sight lines in this study probe a relatively quiescent region of the LMC (Figure~\ref{figure:ha}), the intermediate-velocity absorbing material they trace at $v_{\rm LMC}=\pm100~\kms$ was likely expelled by stellar feedback in the highly active star-forming regions of the LMC (Section~\ref{section:cal_f_absorption}). In this scenario, the outflows become thermalized and have expanded across the disk, making them detectable far from where they were ejected. The LMC has therefore undergone at least one major outflow that coincides with the recent intense stellar activity \citep{2009AJ....138.1243H}.  While the velocity of this outflow is smaller than the escape velocity, most of the outflowing material appears to escape into the MW since there is little evidence of infalling gas  \citep{2002ApJ...569..214H,2007MNRAS.377..687L}. The removal of these outflows is likely aided by tidal interactions between the SMC and MW and ram-pressure stripping with the Galactic halo \citep{2015ApJ...813..110H,2015arXiv150707935S}, implying that the environment of galaxies could play an important role in the gas starvation of LMC-like or dwarf galaxies. 
 
\acknowledgments

We thank Y.-H.~Chu for providing us with the reduced 2D \ha\ spectrum of HD~33133 from \citet{1999AJ....117.1433C} as well as Lister Staveley-Smith and Sungeun Kim for providing us with the ATCA and Parkes telescopes LMC H\textsc{~i} survey datacube. We also acknowledge useful discussions with Tony Wong and Gurtina Besla and comments made by the anonymous referee. This paper includes archived \hi\ LMC data obtained through the Australia Telescope Online Archive (http://atoa.atnf.csiro.au) and the Galactic All Sky Survey Archive (http://www.atnf.csiro.au/research/GASS). This paper also includes archived FUSE data available through the Mikulski Archive for Space Telescopes (MAST: http://archive.stsci.edu). Support for this program  was provided by NASA through through the grant HST-GO-11692 from the Space Telescope Science Institute, which is operated by the Association of Universities for Research in Astronomy, Incorporated, under NASA contract NAS5-26555. K.A.~Barger was supported through NSF Astronomy and Astrophysical Postdoctoral Fellowship award AST~1203059. 

\bibliographystyle{apj} 
\bibliography{References} 

\null\clearpage
\section{Appendix}\label{section:appendix}

\subsection{Wind Nebula surrounding the HD~33133 stellar target}

\begin{figure}
\begin{center}
\includegraphics[scale=0.45,angle=0]{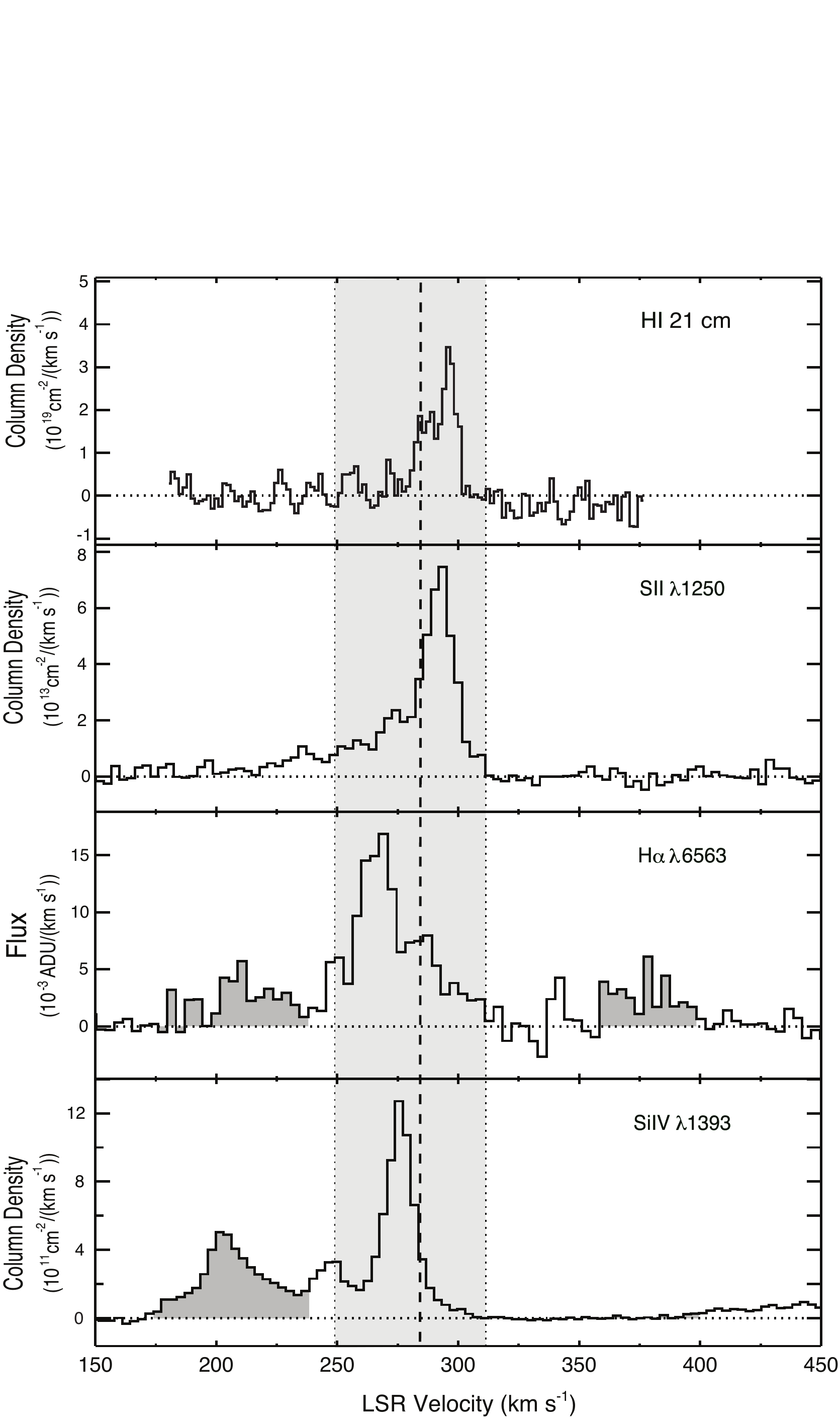}
\end{center}
\figcaption{Comparison of emission- and absorption-lines along the HD~33133 sight line. The H\textsc{~i} emission and the ${\rm S}\textsc{~ii}~\lambda1250$ species trace the cooler gas phase in the top two panels. The combined ATCA and Parkes H\textsc{~i} spectrum traces the small scale structure with an angular resolution of $6\arcmin$ (top panel). The \ha\ emission and ${\rm Si}\textsc{iv}~\lambda1393$ species trace the warmer gas (bottom two panels). We extracted the \ha\ spectrum from a long-slit echelle spectrograph image positioned east-to-west across HD~33133 with a slit width of $1.64\arcsec$; \citet{1999AJ....117.1433C} shows this long-slit image, acquired with the $4-{\rm m}$ telescope and the Tektronic detector at the Cerro Tololo Inter-American Observatory (CTIO),  in Figure~2 (labeled under the alias Br~13). The grey highlighted components span the velocity extent of the approaching $(+175\le\vlsr\le+237~\kms)$ and receding $(+335\le\vlsr\le+400~\kms)$ walls of the H\textsc{~ii} shell surrounding this WR star. To emphasize component structure of this shell, the average \ha\ spectrum is offset slightly offset from HD~33133, spanning $2.6$ to $6.5\arcsec$ to the west. The ${\rm S}\textsc{~ii}~\lambda1250$ and ${\rm Si}\textsc{~iv}~\lambda1393$ apparent column density profiles are only sensitive to the gas in the foreground of HD~33133, whereas the H\textsc{~i} and \ha\ emission is sensitive to both the approaching gas in the foreground and the receding gas in the background. The systemic velocity of the LMC at $\vlsr=+282~\kms$ ($v_{\rm LMC}=0~\kms$) is marked with a black-dashed line. The extent of H\textsc{~i} emission from the LMC disk is highlighted in light grey over the $+250\le\vlsr\le+310~\kms$ ($-32\le v_{\rm LMC}\le+28~\kms$) velocity range as illustrated in Figure~\ref{figure:hi}.
\label{figure:shell}}
\end{figure}

The WR star HD~33133 has created a wind nebula as seen from the \ha\ spatial distribution in the bottom-right panel of Figure~\ref{figure:ha}. The \ha\ emission spectrum shown in Figure~\ref{figure:shell} \citep{1999AJ....117.1433C} also traces this nebula, where the approaching and receding shells are highlighted at $+175\lesssim\vlsr\lesssim+237~\kms$ and $+335\lesssim\vlsr\lesssim+400~\kms$, respectively. Unlike the absorption spectra along this sight line, the emission traces the entire galaxy and its surroundings. Because the ${\rm S}\textsc{~ii}~\lambda1250$ and ${\rm Si}\textsc{~iv}~\lambda1393$ apparent column density profiles---also included in this figure---only trace the absorbing material in front of the star, the receding shell is absent in the stellar spectrum.

The high-ion absorption along the stellar HD~33133 sight line shows distinct, narrow components, whereas the absorption along the AGN CAL~F sight line only consists of broad velocity components. In Section~\ref{section:vel_collisional} and in Figure~\ref{figure:resolution}, we showed that the narrow component structure of the HD~33133 sight line was not due to the differences in the spectral resolution of the COS and STIS instruments. The narrow components in the Si\textsc{~iv} spectrum (and in the spectra of other species) near $\vlsr\approx+200~\kms$ aligns well with the approaching \ha\ shell that surrounds this WR star (see Figures~\ref{figure:ha} and~\ref{figure:shell}). These narrow components indicate that the gas along the HD~33133 sight line is being illuminated by this early-type star and the underlining broad components, similar to those seen along the AGN sight line, suggests that collisional processes also influence this gas. The lack of narrow component structure in the CAL~F spectra implies that this sight line lies outside of any hot stars' direct influence. 

Because of the low signal-to-noise ratio of the C\textsc{~iv} absorption along the HD~33133, we are unable to directly compare the properties of its broad components with those along the CAL~F sight line. However, they both appear to have similar underlining properties with both being highly ionized at $x({\rm H^+})\ge0.72$ in the low-ionization species and likely near unity for the high-ionization species, which are also predominantly collisionally ionized. This suggests that the broad components along these sight lines likely trace the same gas over similar path lengths as these sight lines are projected only $105~{\rm pc}$ apart and are affected by the same large-scale processes. Because the underlying broad components seen in C\textsc{~iv} along the HD~33133 sight line it is blue shifted with respect to the systemic velocity of the LMC. Therefore, this gas must be flowing towards us on the near side of the galaxy. As the CAL~F sight line has an additional red shifted component, that sight line also traces outflowing gas receding from us and the LMC on the far side of the galaxy.

\end{document}